# Correlation between the parameters of the rate equation for simple aftershock sequences: implications for the forecasting of rates and probabilities

Paolo Gasperini and Barbara Lolli

Dipartimento di Fisica
Università di Bologna
Viale Berti-Pichat 8
I-40127 Bologna (Italy),
e-mail: paolo@ibogfs.df.unibo.it, barbara@ibogfs.df.unibo.it

**Abstract**

We analyzed the correlations among the parameters of the Reasenberg and Jones (1989) formula describing the aftershock rate after a mainshock as a function of time and magnitude, on the basis of parameter estimates made in previous works for New Zealand, Italy and California. For all of three datasets we found that the magnitude-independent productivity *a* is significantly correlated with the *b*-value of the Gutenberg-Richter law and, in some cases, with parameters *p* and *c* of the modified Omori's law. In most cases, we also found a significant correlation between *p* and *c* but, different from other previous works, we did not find a significant correlation between *p* and *b*. We argued that the correlation between *a* and *b* can be ascribed to an inappropriate definition of the coefficient of mainshock magnitude as the correlation becomes insignificant if the latter is assumed to be $\alpha \approx 2/3 b$ rather than *b*. This interpretation well agrees with the results of direct $\alpha$ estimates we made, by an epidemic type model (ETAS), from the data of some large Italian sequences. We also verified that assuming $\alpha$ about 2/3 of the average *b* value estimated from Italian sequences occurred in the time interval 1981-1996 improves the ability to predict the behavior of most recent sequences (from 1997 to 2003). The discrepancy with other recent analysis of distributed seismicity, which found $\alpha \approx b$, might be due to a lower fractal dimension of the spatial distribution of shocks belonging to single sequences with respect to the entire seismicity of an area. Our results indicate a partial inadequacy of the original Reasenberg and Jones (1989) formulation when used to forecast the productivity of future sequences. In particular, the aftershock rates and probabilities tend to be overestimated for stronger mainshocks and conversely underestimated for weaker ones. We also inferred that the correlation of *a* with *p* and *c* might be the consequence of the trade-off between the two parameters of the modified Omori's law. In this case the correlation can be removed by renormalizing the time-dependent part of the rate equation. Finally, the absence of correlation between *p* and *b*, observed for all the examined datasets, indicates that such correlation, previously suggested by theoretical inferences and empirical results from some regions, does not represent a common property of aftershock sequences in different part of the world.

## 1. Introduction

Standard Probabilistic Seismic Hazard Assessment (PSHA) methods (e. g. Cornel, 1968) usually completely neglect aftershock occurrence. However, the experience of recent significant earthquakes clearly showed that strong aftershocks are the source of further damages and of an expansion of the area requiring emergency services (Hough and Jones, 1997). Thus a reliable forecasting of the behavior of the aftershock sequences, starting from the first hours since the mainshock, is highly desirable and often is also strongly solicited by civil services, rescue corps and mass-media.

Most current approaches to aftershock-rate modeling refer to scaling laws that were empirically found to fit well the data in different parts of the world, such as the frequency-magnitude Gutenberg and Richter (1944) law (G-R) and the Modified Omori Model (MOM, Utsu, 1971) describing shock rate decay with time. Reasenberg and Jones (1989) proposed a simple comprehensive model, which is still in use, at the U.S Geological Survey, to forecast aftershock rates and probabilities in Southern California and in few other areas of the world. They combined the G-R and the MOM to describe the aftershock occurrence as a non-stationary Poisson process whose rate varies with time $t$ after the mainshock according to

$$\lambda(t) = \frac{10^{a+b(M_m - M_{min})}}{(t+c)^p} \quad (1)$$

where $b$ is the slope of the G-R, $c$ is a small time shift (usually <1 day) that controls the sequence behavior in the first times after the mainshock, $p$ (usually around 1) determines the speed of decay of aftershock rate with time, and $a$ is referred by the authors as the "productivity" of the sequence. $M_m$ is the mainshock magnitude and $M_{min}$ is the lower magnitude threshold of the sequence.

Several other studies, dedicated to the understanding of the aftershock sequence behavior, have been published in the last decades. Among those others, Kagan and Knopoff (1981, 1987) and Ogata (1988) introduced the epidemic principle (ETAS), considering the aftershock generation process as the superposition of simple Omori's decays. The ETAS model has also been used to give short-term probabilistic forecast of seismic activity (Kagan and Knopoff, 1987; Kagan and Jackson, 2000; Console and Murru, 2001). Gross and Kisslinger (1994) demonstrated that the stretched exponential law (Shlesinger and Montroll, 1984; Kisslinger, 1993) may fit the data better than MOM in some cases. Recently, Helmstetter and Sornette (2002) showed that aftershock sequences may follow different regimes (subcritical and supercritical) depending on the "branching ratio", that is the mean aftershock number triggered per event, while Narteau et al (2002, 2003) proposed a "band limited" power law restricted to the time interval between two characteristic times, before and after which the rate decay with time is respectively linear and exponential.

These contributions strongly deepen our understanding on various aspects of the aftershock physics and particularly of their distribution in time, space and magnitude. Also, such models are able to improve the fit with the data of complex sequences in most cases. However, in our present analysis, we concentrate on the Reasenberg and Jones (1989) model because it is very simple and particularly suited to make forecast of the behavior of future sequences. Basically, it allows to compute, representative average values of model parameters from the sequences previously occurred in a certain area. Moreover, although Reasenberg and Jones (1989) model is less accurate and refined than the other ones mentioned above, it has the merit to allow the forecasting of aftershock behavior only basing on the mainshock magnitude while the other models also require the real-time location and sizing of all the aftershocks above a given threshold. Finally, the Reasenberg and Jones (1989) model was applied in different parts of the world with rather uniform procedures since these were established in specific Fortran codes written by Paul Reasenberg (*cluster5* for sequence detection and *aspar3x* for sequence modeling). This allows making inferences whose validity is not restricted to the specific dataset or region considered. Furthermore, since we analyze the results of previously published papers, we avoid the risk of biasing the results by the adoption of *ad hoc* choices for sequence detection and modeling.

## 2. Estimation and meaning of Reasenberg and Jones (1989) model parameters

Program *aspar3x* computes the parameters of the Reasenberg and Jones (1989) model through the maximum likelihood estimate (MLE) technique, from the data of real sequences. While the *b*-value of the G-R law is determined by fitting the distribution of aftershock magnitudes independently of time, using the procedure firstly developed by Aki (1965), the other parameters are estimated maximizing, with respect to the values of parameters *p* and *c*, the likelihood (*i.e.* Ogata, 1983) of the MOM equation

$$\lambda(t) = \frac{K}{(t+c)^p} \tag{2}$$

where *K* is computed by constraining the rate integral from *S* (initial time) to *T* (end time) to be equal to the total number *N* of observed shocks with $M \geq M_{min}$

$$N = \int_S^T \lambda(t)dt = K \int_S^T (t+c)^{-p} dt \tag{3}$$

Thus, sequence productivity *a* of eq. (1) is simply obtained as

$$a = \text{Log}_{10}(K) - b(M_m - M_{min}) = \text{Log}_{10}(N) - b(M_m - M_{min}) - \text{Log}_{10}\left[\int_S^T (t+c)^{-p} dt\right] \quad (4)$$

where the integral in last term can be easily computed as

$$\int_S^T (t+c)^{-p} dt = \begin{cases} \dfrac{(T+c)^{1-p} - (S+c)^{1-p}}{1-p} &, p \neq 1 \\ \ln\left(\dfrac{T+c}{S+c}\right) &, p = 1 \end{cases} \quad (5)$$

In expressions (1) and (4), $bM_{min}$ accounts for the dependence of the number of shocks on the minimum-magnitude threshold, according to the G-R law. The term $a+bM_m$ corresponds to the G-R intercept of the considered dataset and can be seen as the global aftershock productivity of the given sequence. In such term, $a$ represents the magnitude independent productivity, maybe related to the local physical properties of the area where the sequence occurs, while $bM_m$ is the productivity factor related to mainshock magnitude.

Differently from $p$ and $c$, that only depend on the time distribution of shocks, and $b$, only depending on shock magnitudes, parameter $a$ is a function of both time and magnitude distributions. We can consider it as the sum of two factors: a "true" magnitude independent productivity and a time-dependence normalizing factor. In principle, the separation of the two factors could be possible when $p>1$ as, in this case, the total number of shocks generated by the mainshock, in the time interval t=[0,∞], is finite and can be computed as

$$N_{tot} = \int_0^\infty \lambda(t) dt = K \int_0^\infty (t+c)^{-p} dt = N \frac{\int_0^\infty (t+c)^{-p} dt}{\int_S^T (t+c)^{-p} dt} \quad (6)$$

Then the true productivity could be defined as

$$a_0 = \text{Log}_{10}(N_{tot}) - b(M_m - M_{min}) = a + \text{Log}_{10}\left[\int_0^\infty (t+c)^{-p} dt\right] \quad (7)$$

Consequently the rate equation would become

$$\lambda(t) = \frac{10^{a_0 + b(M_m - M_{min})}}{(t+c)^p \int_0^\infty (t'+c)^{-p} dt'} \quad (8)$$

Such equation also makes clearer the dimensional consistency of the formulation as the numerator is a pure number and the denominator a time. However such separation is not possible in general, as if $p \leq 1$, for any $M_{min}$, both $N_{tot}$ and $a_0$ tend to ∞ (albeit $a$ is finite).

Actually, the divergence of $N_{tot}$ is likely a mathematical paradox rather than a physical reality because physical arguments and empirical evidences suggest that the total number of aftershocks of a real sequence must be finite. In fact, Narteau et al. (2002) have recently shown that a physically grounded model, with a power law decay limited in time, fits the data better then the simple MOM.

They showed that, for times longer than a critical value $t_2$ (some tens or hundreds of days, depending on the *p* value), real sequences exhibit a faster exponential decay. This implies that, even if the rate for $t<t_2$ is well described by a MOM with $p\leq 1$, the total number of aftershocks generated by a mainshock could be finite. For the sake of an approximate estimation of the true productivity of the sequences, we could limit the computation of the normalization integral of eq. (7) and (8) to a finite time window, definitely larger than $t_2$ (usually few years), after which the sequence rate and even the remainder of the integral are assumed to be negligible.

It has been noted already (Helmstetter, 2003) that taking *b* as the linear coefficient of $M_m$ is only an hypothesis, also assumed by many other studies (Kagan and Knopoff, 1987; Davis and Frolich, 1991; Console and Murru, 2001; Felzer et al., 2002), that lacks of a clear justification. It can be heuristically deduced from the assumption that the mainshock and the aftershocks belong to the same population and follow the same G-R law. Although this might appear reasonable, there are not strict physical arguments or univocal empirical evidences proving that the productivity of a mainshock scales with magnitude with a coefficient equal to *b*. Helmstetter et al (in press) suggest that such coefficient is strictly related to the fractal dimension *D* of hypocenters distribution. From physical considerations on the nature of the triggering stress they inferred a relation $\alpha \approx 0.5D$ which, on the basis of the estimate $D\approx 2$, made for a network of faults in California, brings to $\alpha\approx 1$ and thus $\alpha\approx b$.

Ogata (1988) and subsequent papers made a different assumption with respect to Reasenberg and Jones (1989): they still took a common coefficient $\alpha$ for both $M_m$ and $M_{min}$ but considered it as a different parameter to be estimated from data independently of *b*. Recently, Console *et al.* (2003), applying an ETAS model on the entire Italian seismicity, found a value of mainshock productivity coefficient about 0.5*b*. Furthermore, Helmstetter (2003), by a stacking of the catalogue of Southern California earthquakes, computed a value around 0.8, significantly lower than the average *b*-value although the subsequent paper (Helmstetter et al., in press), making some different choices in the clustering algorithm, came to the conclusion that mainshock productivity coefficient is 1.05±0.05 which also agrees with the independent result obtained by Feltzer et al. (2004). It is to be noted however that the latter approaches capture the features of a process – the entire seismicity of a extended area over a time interval of some tens of years – essentially different from the one considered in this paper: the occurrence of aftershocks in the months after the main shocks and along well defined faults. As well, they are all almost unusable for forecasting as they would require the detection and sizing, in near real-time, of all of the shocks (above a certain magnitude threshold) occurring in the considered area, which can hardly be performed by seismological observatories during a very productive seismic sequence.

An empirical approach to settle the question regarding the correct value of mainshock magnitude coefficient derives from the following consideration: if the magnitude productivity term $bM_m$ of the Reasenberg and Jones (1989) model well describes the dependence of the number of aftershocks on mainshock magnitude we would have to expect that it is not correlated with the magnitude-independent productivity $a$. If instead such a interplay of the two term is effective then it can be the symptom of an inadequate formulation of the two terms. This argument addresses in a simple empirical way a more general problem concerning the search for an orthogonal set of model parameters (Cox and Reid, 1987). If we exactly know the form of a model with a given set of parameters and these are not orthogonal it is possible to deduce an orthogonal set that is more efficient in estimating the parameters from the data by MLE (*i.e.* Yang and Xie, 2003). When instead the physical grounding, like in this case, is weak or lacking we cannot establish if the parameters actually represent independent effects rather than combinations of factors and the analysis of correlation among the parameters can be used to find an optimal parameter set.

We want to point out that the choice of model parameters representing truly independent physical properties of the aftershock occurrence process is crucial when their estimates are averaged to compute *a-priori* values to be used for the forecasting of future sequences as done for example by Reasenberg and Jones (1989), Eberhart-Phillips (1998) and Lolli and Gasperini (2003). The use of hybrid variables, depending on the combination of different factors, reduces the accuracy of the estimates and at the same time the reliability of the forecasting model. For example parameter $K$ of the simple modified Omori law is completely unusable for the forecasting as it depends on the lower magnitude threshold and on the mainshock magnitude. Thus estimates of $K$ from different sequences are not comparable among each other and the average of $K$ values is not significant for the forecasting. In this sense, parameter $a$ of the Reasenberg and Jones (1989) model is instead, at least partially, "debiased" from these effects and can be tentatively averaged to predict the productivity of future sequences.

**3. Testing the correlation among aftershock model parameters**

The correlation among the parameters of the scaling law describing the aftershock behavior as a function of time, space and magnitude has been investigated in many theoretical papers (Utsu, 1961; Mikumo and Miyagate, 1979; Main, 1992). Guo and Ogata (1995, 1997) studied parameter empirical correlation basing on the data of interplate and intraplate Japanese sequences. They found a clear correlation between some couples of parameters but in some cases with opposite sign for intraplate and interplate earthquakes. For example the correlation between $p$ and $b$ is positive for intraplate earthquakes and negative for interplate ones. They suggested that these differences might reflect some intrinsically different features in the rupturing process between earthquakes within a plate and on a margin between two plates.

In our present analysis we mainly consider the estimates of the Reasenberg and Jones (1989) model parameters made by Lolli and Gasperini (2003) for Italy and by Eberhart-Phillips (1998) for New Zealand as these works list the parameter values for the different sequences (Datasets 1 and 2 respectively in Appendix A). Unfortunately this is not the case of the original work for California by Reasenberg and Jones (1989). However, we have received a list of sequence parameters (Dataset 3) from one of the authors (Jones, personal communication) that allows us to make some computations even on such data.

The *aspar3x* program was used by Lolli and Gasperini (2003) and Eberhart-Phillips (1998) to homogeneously estimate the parameters of the rate formula. Moreover, the Reasenberg (1985) clustering program (*cluster5*) was used, by both works, to associate events into aftershock sequences. Lolli and Gasperini (2003) also employed a simpler clustering algorithm, similar to the one proposed by Gardner and Knopoff (1974), based on a space-time window varying with mainshock magnitude. A comparison made by Lolli and Gasperini (2003) among corresponding sequences detected by different methods showed that the influence of the clustering algorithm on estimated parameters is almost negligible. The procedure followed by Reasenberg and Jones (1989) was instead slightly different as they fixed $c=0.05$ for all the sequences. This prevents correlation analyses involving such parameter for California dataset.

We evaluated the existence of linear correlation among couples of variables by computing the linear correlation coefficient $r$ and estimating its statistical significance (Davis, 1986, pp. 67). We compared the $H_0$ hypothesis that the linear correlation coefficient $r$ is equal to 0 versus the alternative that it is different from 0 using the Student-t statistics

$$t = \sqrt{\frac{r^2(n-2)}{1-r^2}} \qquad (12)$$

If the value of $t$ is greater than the critical threshold (found in common statistical tables) corresponding to the significance level s.l.=0.05 for the given degrees of freedom $n$-2, the $H_0$ hypothesis can be confidently rejected and thus the existence of correlation can be asserted. Conversely, if $t$ is smaller than the threshold the correlation can be excluded. The s.l. shown in following tables indicate the probability that a larger $t$ may occur by chance if $r=0$.

## 3.1 Correlation between *a* and *b*

A significant negative correlation between parameters *a* and *b* is evidenced in Table 1 for the joint dataset including the sequences from Italy, New Zealand and California as well as for the three regions separately (Fig 1). Similar correlations can also be found between *a* and $bM_m$ (Fig. 2). The estimated values of the regression coefficient are quite similar for the different datasets (–0.34±0.14 for New Zealand, –0.37±0.13 for Italy and –0.35±0.06 for California) as well as the regression intercepts that are close to 0 for Italy and California and about 0.33 for New Zealand. The joint dataset including data from all of the three regions gives a slightly different coefficient (–0.30±0.05) and an intercept of 0.22±0.25. On the contrary, the absence of correlation between *a* and $bM_{min}$ (Table 1) in most cases confirms that the latter term quite well accounts for the cutting of earthquakes below $M_{min}$, according to the G-R law.

As argued above, the observed correlation between *a* and $bM_m$ might be the symptom of the inappropriateness of assuming the coefficient of mainshock magnitude equal to *b*. Our analysis seems to give a preference to an average mainshock magnitude coefficient of about *2/3b*. We can rewrite equation (1) as

$$\lambda(t) = \frac{10^{a_1 + \alpha M_m - bM_{min}}}{(t+c)^p} \tag{13}$$

where we can assume

$$\begin{aligned} \alpha &= 0.65b \\ a_1 &= a + 0.35bM_m \end{aligned} \tag{14}$$

Table 1 gives the confirmation that now $a_1$ is uncorrelated to $\alpha M_m$ for all the three datasets. As well, in all cases, excluding the joint set with the data of the three regions together, $a_1$ is not correlated even to *b* (Fig. 3). Note that the peculiarity of the California dataset, showing a clear negative significant correlation between *a* and $bM_{min}$, is very likely an artifacts of the combination of the correlation found between *a* and $bM_m$ and of the choice of taking $M_{min}=M_m-3$ for all the sequences, made by Reasenberg and Jones (1989).

Eq. (13) is not directly usable to model single sequences as parameters $a_1$ and $\alpha$ cannot be estimated simultaneously. However, we can assume $\alpha \cong 0.65b$ as an average property of all sequences in different regions of the world and estimate a different value of $a_1$ for each sequence. We must underline that eq. (13), when used to forecast the behavior of future sequences, implies a definitely different aftershock-rate dependence on mainshock magnitude (often the only reliable information available on the current sequence in the hours immediately after a mainshock) with respect to the original formulation. In particular it predicts lower rates than Reasenberg and Jones (1989) for stronger mainshocks and higher rates for weaker ones. It might be argued that this could be due to a bias in the selection of the sequences. Particularly, the sequences related with low magnitude mainshocks could be included in the datasets only if they had an unusually large number of aftershocks and thus a particularly high productivity. However this can be excluded, at a first order at least, since a significant linear correlation between $a$ and $bM_m$ can even be found (although with a slightly lower coefficient) for the dataset limited to the sequences with mainshock magnitude equal or larger than 6.0 (Table 1) for which the sequence catalogue is likely to be complete.

Another possible objection to the interpretation of the observed correlation between $a$ and $bM_m$ given in terms of the inadequacy of the formulation, could come from the role played by parameter estimation errors. As $a$ is computed as a function of $b$, the correlation can be artificially induced, at least in part, by random errors on $b$. This effect is not easy to evaluate, however if it is significant we should have to observe a similar correlation also between $a$ and $bM_{min}$ and we should have to correct even the coefficient of $M_{min}$ to remove the correlation. As such correlation is not effective for both Italy and New Zealand (and it is due to aftershock selection rules for California) we can deduce that estimation errors are not likely to be the main source of the observed correlation between $a$ and $bM_m$.

The regional averages of parameter $a_1$, reported in Table 2 (0.02±0.06 for California, –0.12±0.13 for Italy and 0.38±0.16 for New Zealand), approximately corresponds to the intercept terms of the regressions between $a$ and $bM_m$ for the same areas. We can hypothesize that the significant difference between New Zealand and other two regions reflects the local characteristics of the seismogenic area but, at the same time, that it can be artificially induced by the different calibration of the magnitude scales. We must note, in fact, that the magnitude scale adopted for New Zealand is Mw (Eberhart-Phillips, 1998), while Ml (calibrated with real and synthetic Wood-Anderson instruments) was used for Italy (Gasperini, 2002; Lolli and Gasperini, 2003) and for most sequences of California (Reasenberg and Jones, 1989). We made an attempt to make the magnitude of the New Zealand catalogue homogeneous with the other two, using the empirical relation found for the Italian catalogue by Gasperini and Ferrari (2000)

$$\mathrm{Log}_{10}(M_0) = 1.22(\pm 0.08)M_l + 17.70(\pm 0.40) \quad (R^2 = 0.87) \quad (15)$$

This formula lies about in the middle of the distribution of empirical laws estimated in various areas (Thacher and Hanks, 1973; Backun, 1984; Johnston, 1996; Backun and Lindt, 1977; Bolt and Herraiz, 1983; Chavez and Priestley, 1985) and represents a reasonable compromise among them. We did not use the specific relation available for New Zealand by Dowrick and Rhoades (1998) as this is poorly constrained by the data ($R^2$=0.65), probably because it is based on Ml estimates mostly made not using true or synthetic Wood-Anderson instruments.

When $M_m$, $M_{min}$ and $b$ values for New Zealand dataset are converted from Mw to Ml by inversely using equation (15), the average value of $a_1$ becomes -0.03±0.16 (Table 2) as well as the $a_1$ vs $bM_m$ regression, for both the joint datasets (Italy plus New Zealand and all the three regions together), gives coefficients of about –0.35 and intercepts close to 0 (Table 1), very similar to the values found for separate regions. Although the zero-close averages of parameter $a_1$ for the all regions may only be due to the chance, we must note that our analysis indicates a surprising homogeneity of the average behavior of seismic sequences in different parts of the world.

Owing to the high dispersion of the data, it can be argued that regression coefficients and correlation significances might be biased by outliers and by a deviation from normality of residuals distribution. For testing this possibility we applied a bootstrap resampling technique (Efron and Tbishirani, 1986; Hall, 1992) to the $a$ vs $bM_m$ regression. We extracted 5000 random samples (with repetition) from the different datasets and computed means and medians of regression parameters and statistics. Table 2 shows the results of bootstrap test for single regions and for joint datasets. In all cases both the means and medians give values very close to those estimated by standard least-squares regression thus allowing to exclude the existence of a bias induced by outliers or by deviations from normality.

Another effect of the parameter transformation from $a$ to $a_1$, shown in Table 3, is the reduction of the standard deviation (of 15% on average) both for the separate regional set and for the joint one. This seems to indicate that the new parameter definition better corresponds to a common property (the true magnitude-independent productivity) of the sequences.

**3.2 Correlation between *a* and MOM parameters**

Table 4 shows that parameters $a$ and $a_1$ are significantly and positively correlated with $p$ for both the joint datasets including sequences from Italy and New Zealand (Fig 4) and from all the three regions together. A similar correlation can also be found, for the former dataset, between parameters $a$ and $c$ (Fig.5). We also report in Table 4 the results of correlation analyses (which are also significant) with respect to Log $c$ (Fig.6) because the latter describes the parameter distribution better than $c$ itself (Lolli and Gasperini, 2003). Weaker correlations can be found when considering Italy and New Zealand datasets independently. On the contrary, for California, the correlation between $a$ and $p$ is totally insignificant, and between $a$ and both $c$ and Log $c$ is not meaningful because $c=0.05$ is assumed for all the sequences of this dataset.

Although relatively weak, the correlations found can be tentatively ascribed to the effect of the implicit inclusion in parameters $a$ and $a_1$ of the time-dependence normalization integral. In fact, if we remove the term

$$Int = -\text{Log}_{10}\left[\int_0^T (t+c)^{-p} dt\right] \qquad (16)$$

from both *a* and $a_1$ (in Table 4, parameters $a_0=a-Int$ and $a_2=a_1-Int$), assuming for all sequences *T*=1096 days (3 years) as upper limit after which the remainder of the time integral is considered to be negligible (see above), the correlation becomes not significant in all but one case with respect either to *p*, *c* and Log *c*. The average value of $a_2$ is about 1.0 for all the separate dataset as well as for the joint one (Table 3). Moreover even for this reformulation the standard deviations of transformed parameters are lower for Italy and New Zealand and to a smaller extent for the joint dataset (Table 2). Conversely the variances slightly increase for California where, as noted above, the procedure followed to estimate Omori's law parameters was significantly different.

With this modification the rate formula would become

$$\lambda(t) = \frac{10^{a_2 + 0.65bMm - bM\min}}{(t+c)^p \int_0^T (t'+c)^{-p} dt'} \qquad (17)$$

Due to the approximate procedure we have used to estimate the normalization integral and the weakness of the correlation found, equation (17) is far from being significant and reliable enough to be proposed as a possible rate formula for the actual modeling and forecasting of simple sequences. However the fact that $a_2$ is now independent of the other parameters (*b*, *p*, and *c* in Tables 1 and 4) suggests that its regionally or locally averaged values could be more suitable than those computed from $a_1$ (or *a*) to predict the behavior of future sequences occurring in the same areas.

### 3.3 Correlation among other parameters

In Table 5 we report the results of correlation analyses among *b*, *p* and *c*. We find a weak negative correlation (r=–0.22) between *b* and *p*, only for the joint dataset of all the three regions together, but not for any of the separate datasets (Fig. 7) and for Italy plus New Zealand. This is somehow at odds with heuristic considerations made by Utsu (1961) and even with empirical estimates by Guo and Ogata (1995, 1997), which found instead clear correlations (with opposite signs), for a regional subdivision of Japanese sequences (interplate vs intraplate). In our case, the correlation found for the joint set seems instead to be the effect of the artificial combination of non-homogeneous data rather than a clear property of the sequences in different regions. The correlation between *b* and *c* (or Log *c*) is instead not significant in all cases.

For Italy and particularly for New Zealand we find a clear positive correlation between *p* and *c* (Fig. 8) or Log *c* (Fig. 9). This is related to the interplay of the two parameters within the MLE procedure. In fact, it is easy to understand how an high *c* value (maybe induced by the incompleteness of the data in the first times after the mainshock) has the effect of slowing the sequence decay for times $t<c$, and this in turn can be counterbalanced by an increase of the *p*-value. This interplay is clearly revealed by Fig. 10, where we plotted the contours of the log-likelihood function, for one of the sequences (B17) analyzed by Lolli and Gasperini (2003). It is easy to understand how the location of the absolute maximum (marked with an asterisks) is only barely constrained along the direction corresponding to the major axis of the highest contour level and thus could vary along such direction due to the effect of data uncertainties and incompleteness.

A more deepened discussion of this argument is postponed to a future study as it must involve the analysis of the physical significance of *p* and *c*, within the framework of more physically grounded models of aftershock occurrence such as those proposed by Kisslinger (1993) and Narteau et al. (2002, 2003).

## 4. Validation of the modified rate formula

To compare the forecasting efficiency of the modified formulation defined by eq. (13) and (14) with respect to the original one by Reasenberg and Jones (1989), we analyzed a set of sequences, occurred in Italy from 1997 to 2003, not considered by Lolli and Gasperini (2003) to estimate the average values of the rate formula parameters. We used data taken from the on-line seismic bulletin of the *Istituto Nazionale di Geofisica e Vulcanologia* (INGV), relocated with the same procedures applied to the catalog (CPTI Working Group, 2001) used by Lolli and Gasperini (2003) and with magnitude revalued according to Gasperini (2002). We followed the same method applied in our previous paper to detect the sequences (using both *cluster5* code and a simpler algorithm with time and space window varying with mainshock magnitude) and to evaluate their goodness of fit with the Reasenberg and Jones (1989) model (by $\chi^2$ and Kolmorogov-Smirnov tests computed by the *aspar3x* program). As well, we selected only the sequences with mainshock magnitude not smaller than 4.3 and including at least 20 shocks. The list of detected sequences, reported in Table 6, does not include the long and particularly productive macro-sequence occurred, from September 1997 to June 1998, in the Umbria-Marche region as its complexity (5 strong shocks in the range of Ml magnitude from 5.4 to 5.8) prevented a good fit with the simple MOM.

We simulate the forecasting of the Italian sequences from 1997 to 2003, for both the original Reasenberg and Jones (1989) model and the modified one, using the average parameters previously estimated by Lolli and Gasperini (2003). For the original Reasenberg and Jones (1989) model, as suggested by Lolli and Gasperini (2003), we use the averages (arithmetic mean, weighted mean and median) of parameters *a* and *b* relative to only the 20 sequences occurred from 1981 to 1996 (sequences from B17 to A30 of Dataset 1) when the catalog was more reliable. For the modified model of eq. (13), we estimate the averages of parameters $a_1$ and $\alpha$ by eq. (14) for the same set of sequences (see Table 7). In both cases we used averages of parameters $p$ and $c$ as reported by Lolli and Gasperini (2003) (also in Table 7). We compared the relative efficiency of the two formulations by computing the ratio of their likelihoods with the new sequences.

The log-likelihood function for the Reasenberg and Jones (1989) model is given by (*i.e.* Ogata, 1983; Lolli and Gasperini, 2003)

$$\ell_{RJ}(a,b,p,c) = N[a + b(M_m - M_{min})]\ln 10 - p\sum_{i=1}^{N}\ln(t_i + c) - 10^{a+b(M_m - M_{min})}\int_{S}^{T}(t+c)^{-p}dt \quad (18)$$

where $t_i$ and $N$ are respectively the occurrence times and the total number of shocks in each sequence within the time range from S to T. For the modified model of eq. (13) the log-likelihood function is given by

$$\ell_{GL}(a_1,b_1,p,c) = N[a_1 + \alpha M_m - bM_{min}]\ln 10 - p\sum_{i=1}^{N}\ln(t_i + c) - 10^{a_1+\alpha M_m - bM_{min}}\int_{S}^{T}(t+c)^{-p}dt \quad (19)$$

so that the log-likelihood ratio becomes

$$LR = \ell_{RJ} - \ell_{GL} = N[a - a_1 + M_m(b - \alpha)]\ln 10 - [10^{a+b(M_m - M_{min})} - 10^{a_1+\alpha M_m - bM_{min}}]\int_{S}^{T}(t+c)^{-p}dt \quad (20)$$

If *LR* is positive, then the Reasenberg and Jones (1989) model works better than the modified one. Conversely the latter overperforms the former when *LR* is negative.

From Table 6 we can see how *LR* values are negative (highlighted with bold type) for the majority of the sequences (11 over 17). As well the cumulative *LR* values (sum of all sequence *LR* values) are negative in all cases. These results confirm the stability with time of our findings in Italy and indicate that the modified model might significantly improve the ability to forecast future sequences behavior.

## 5. Application of the modified formula to ETAS processes

The modified rate formula (13) can be easily applied to more complex processes of epidemic type (Ogata, 1988) where all the shocks of the sequence, and not only the mainshock, are assumed to induce further aftershocks. Under the hypothesis that the parameters of eq. (13) are the same for all "mother shocks" (generating aftershocks), the global aftershock rate equation becomes

$$\lambda(t) = \sum_{j=1}^{N_m(t)} \frac{10^{a_1 + \alpha M_j - bM_{min}}}{(t - u_j + c)^p} \quad (21)$$

where $N_m(t)$ is the number of mother shocks (including mainshock) occurred before time $t$; $u_j$ and $M_j$ are respectively the occurrence time and the magnitude of the $j$-th mother shock. We can note that the value of $\alpha$ is now constrained by the magnitude distribution of mother shocks and hence can be estimated simultaneously to $a_1$. On the other hand, the role of parameters $a_1$, $p$, and $c$ in eq. (21) is not exactly the same as in eq. (13), thus we can expect that their MLE are slightly different from the ones obtained by the simple MOM. In this simplified approach we completely neglect the spatial distribution of shocks that thus are all considered to occur at the same place of the mainshock. Although this assumption might appear a defect, it however reduces the risk that the smaller mother-shocks are favoured in predicting the location of triggered seismicity with respect to the stronger ones, and thus that their contribution to triggering might be overestimated (see discussion in Helmstetter et al, in press).

According to the procedure described by Ogata (1983), the log-likelihood function corresponding to eq. (17) is given by

$$\ell(p,c,a_1,\alpha) = N(a_1 - bM_{min})\ln 10 + \sum_{i=1}^{N} \ln\left[\sum_{j=1}^{N_m(t_i)} \frac{10^{\alpha M_j}}{(t_i - u_j + c)^p}\right] - 10^{a_1 - bM_{min}} \sum_{j=1}^{N_m(t_N)} 10^{\alpha M_j} \int_0^{T-u_j} (t+c)^{-p} dt$$

(22)

where again, $N_m(t_i)$ and $N_m(t_N)$ are the number of mother shocks occurring before the shock at time $t_i$ and before the last shock of the sequence respectively. We maximized equation (22), for the data of some Italian sequences detected by Lolli and Gasperini (2003) using the Fortran routine BCONG/DBCONG of the IMSL Math library (Visual Numerics, 1997). Such code uses a quasi-Newton method (Dennis and Schnabel, 1983) and an active set strategy (Gil and Murray, 1976) to solve optimization problems subject to simple bounds.

In order to reliably constrain parameter $\alpha$, we only considered sequences with at least 200 shocks above $M_{min}$. Moreover, we applied the same procedure to some complex sequences that Lolli and Gasperini (2003) were not able to model due the bad fit or that they had to subdivide into subsequences as well as to the long and very productive sequence occurred from September 1997 to June 1998 in the Umbria-Marche region.

The results are shown in Table 8, where the parameters of the Reasenberg and Jones (1989) model, estimated by Lolli and Gasperini (2003) for the same sequences, are also reported when available. In general we can note that, according to previous findings (Guo and Ogata, 1997), $p$ is slightly larger for the ETAS model than for the simple MOM. For only one sequence (B35), characterized by a very small mainshock magnitude (Ml=4.3), the $p$ value for the ETAS model results smaller than 1.0.

On the other hand we can also observe that estimated $\alpha$ values are in most cases significantly lower than $b$. The arithmetic averages are respectively $\bar{\alpha} = 0.79 \pm 0.07$ and $\bar{b} = 1.03 \pm 0.11$ ($\bar{\alpha}/\bar{b} = 0.77$). This difference becomes more evident when considering instead the averages naturally weighted with the inverse of the squared errors that are respectively $\bar{\alpha} = 0.62 \pm 0.02$ and $\bar{b} = 0.89 \pm 0.03$ ($\bar{\alpha}/\bar{b} = 0.69$). These results indicate that, for the analyzed sequences, a coefficient of $M_m$ definitely smaller than $b$ seems to better account for the magnitude dependence of mother-shock productivity.

## 5. Conclusions

From a correlation analysis of parameter estimates, made by different authors in three regions of the world, we deduced a modified form (eq. 13) of the Reasenberg and Jones (1989) time-magnitude distribution of simple aftershock sequences where the parameters are almost independent of each other and thus are more likely to represent true physical properties of the aftershock occurrence process. In particular an average value of magnitude productivity coefficient $\alpha$ about $2/3b$ as well as an average value of magnitude independent productivity $a_1$ close to 0 seem to represent common characteristics of aftershock sequences in different regions.

The proposed modification have a significant impact on forecasting since it implies a definitely different dependence of aftershock rate as a function of mainshock magnitude with respect to one predicted by the original model. Namely, for stronger mainshocks, aftershock rates are lower for the modified model than for the original one and conversely they are higher for weaker mainshocks. We verified that the modified formula improves the ability to forecast aftershocks behavior in the first hours or days after the mainshock, before specific information on the ongoing sequence are available, by comparing the log-likelihood functions of the original and modified formulation with a different set of Italian sequences.

We also introduced in the formulation of eq. (13) the epidemic principle (Ogata, 1988) where each shock of the sequence is the source of further aftershocks. Differently from the simple model with a single mainshock, the rate equation (21) allows to estimating at the same time both parameters $a_1$ and $\alpha$ from the data of a single sequence as the latter is now constrained by the magnitude distribution of mother shocks. The results of the application of this epidemic model to a set of Italian sequences well agrees with our previous findings as it gives averages of parameter $\alpha$ definitely lower than $b$ by a factor ranging from 0.7 to 0.8.

Values of $\alpha<b$ have been estimated already by ETAS modeling of the entire Italian seismicity (Console et al., 2003) and by a stacking of triggered events in Southern California (Helmstetter, 2003). These works found $\alpha=0.5b$ and $\alpha=0.8$ respectively although a more recent paper (Helmstetter et al, in press), using a slightly modified procedure, have got $\alpha=1.0$ for the same dataset of Helmstetter (2003). The latter estimate also agrees with the work by Felzer et al. (2004) which found a value close to 1.0 being compatible with the seismicity of California. The discrepancy between our results and such works could be due to the fact that we capture the properties of a somehow different phenomenon, as we consider relatively short aftershock sequences, well fitting a simple MOM decay, rather than the entire seismicity of a region over a long time interval. In fact, the aftershocks belonging to simple sequences usually occur along single faults, then the fractal dimension $D$ of their hypocentral distribution is likely to be significantly lower than the value expected for the long-term seismicity of a region. Together with the hypothesis that $\alpha \approx D/2$ this would bring to a value of the coefficient of mainshock magnitude for the triggered seismicity along single faults significantly lower than 1.0 found by Helmstetter et al. (in press) and Felzer (2004). As well, the significantly higher $\alpha$ value we observed in ETAS modeling of Italian sequences, with respect to the work of Console et al. (2003), could be explained by our choice of neglecting the spatial distribution of shocks: this in fact reduces the risk that the smaller mother shocks are favoured in predicting the location of triggered seismicity with respect to the stronger ones and thus that their productivity is overestimated.

On the other hand, differently to other models which requires the real-time location of all shocks above a minimum threshold, but similarly to the Reasenberg and Jones (1989) one, our modified formula can be immediately used to forecast aftershock rates and probabilities in the first times after a mainshock, only on the basis of mainshock magnitude. This can be done through eq. (13) taking $\alpha=0.65b$ and the averages of $a_1$ reported in Table 3 for Italy, New-Zealand and California.

In agreement with Kisslinger and Jones (1991), who analyzed most of the California sequences of our dataset 3, we found the absence of correlation between $p$ and $b$ in most cases. This contrasts with heuristic considerations made by Utsu (1961) as well as with empirical estimates made for Japan by Guo and Ogata (1995, 1997). Although the point remains controversial, we can confidently assert that such correlation does not represent a general property of aftershock sequences in different parts of the world.

The weak but significant positive correlation found between $p$ and $c$ in most cases might indicate a bias in the estimates of parameter $p$ due to its interplay with parameter $c$. A higher $c$ value corresponds, in fact, to a slower decay in the first times after the mainshocks that can be counterbalanced by a higher *p*-value and vice versa. This interplay could even be the source of the correlation found between MOM parameters and magnitude independent productivity *a*. In fact, while different combinations of *p* and *c* may give, in many cases, rather similar rate decays within the observation time window [*S*,*T*], the ratio between the number shocks *N* actually observed in the same time window and the total number of shock $N_{tot}$ may vary considerably as a function of *p*. Even considering the incompleteness of the aftershock datasets (particularly in the first times after a strong mainshock), that might influence the estimate of parameter *c*, this variability can significantly affect the estimated value of *a*. This interpretation is confirmed by the absence of correlation between *a* and *p* for the California dataset where *c* is kept fixed to 0.05 and thus cannot influence the estimate of *p*.

The trade-off between *p* and *c* might also indicate a general inadequacy of the MOM in describing the real properties of simple sequences that however requires further deepening in the light of more physical approaches to the definition of aftershock decay process recently developed. In this view, the renormalization of the time dependent part of the rate equation could be useful to separate it from the magnitude dependent part and to make estimates of the sequence productivity that are completely independent of the time decay law.

**Acknowledgements**

We thank Agnès Helmstetter and Paul Reasenberg for a thoughtful review of an earlier version of the manuscript.

**Figure captions**

Figure 1 – Scatter plots and regression lines of *a* versus *b* for the three regions. The correlation is significant in all cases with s.l. <0.01.

Figure 2 – Scatter plots and regression lines of *a* versus $bM_m$ for the three regions. The correlation is significant in all cases with s.l.<0.05.

Figure 3 – Scatter plots and regression lines of $a_1$ versus *b* for the three regions. The correlation is not significant in all cases with s.l.=0.15, 0.35 and 0.10 respectively.

Figure 4 – Scatter plots and regression lines of *a* and $a_1$ versus *p* for the joint dataset of Italy (squares) and New Zealand (diamonds). The correlation is significant in both cases with s.l.=0.02.

Figure 5 – Scatter plots and regression line of *a* and $a_1$ versus *c* for the joint dataset of Italy (squares) and New Zealand (diamonds). The correlation with *a* is significant with s.l.=0.02, while the correlation with $a_1$ is not significant with s.l.=0.06.

Figure 6 – Scatter plots and regression lines of *a* and $a_1$ versus Log *c* for the joint dataset of Italy (squares) and New Zealand (diamonds). The correlation is significant in both cases with s.l.<0.05.

Figure 7 – Scatter plots and regression lines of *b* versus *p* for the three regions. The correlation is not significant in all cases with s.l.=0.31, 0.66 and 0.09 respectively.

Figure 8 – Scatter plots and regression lines of *p* versus *c* for Italy and New Zealand. The correlation is significant in both cases with s.l.=0.02.

Figure 9 – Scatter plots and regression lines of *p* versus Log *c* for Italy and New Zealand. The correlation is significant in both cases with s.l.<0.05.

Figure 10 – Contour lines of MOM log-likelihood as a function of parameters *p* and *c* for the Italian sequence B17 detected by Lolli and Gasperini (2003). The asterisk indicates the exact location of the absolute maximum (*p*=1.27, *c*=0.39).

**Figure 1**

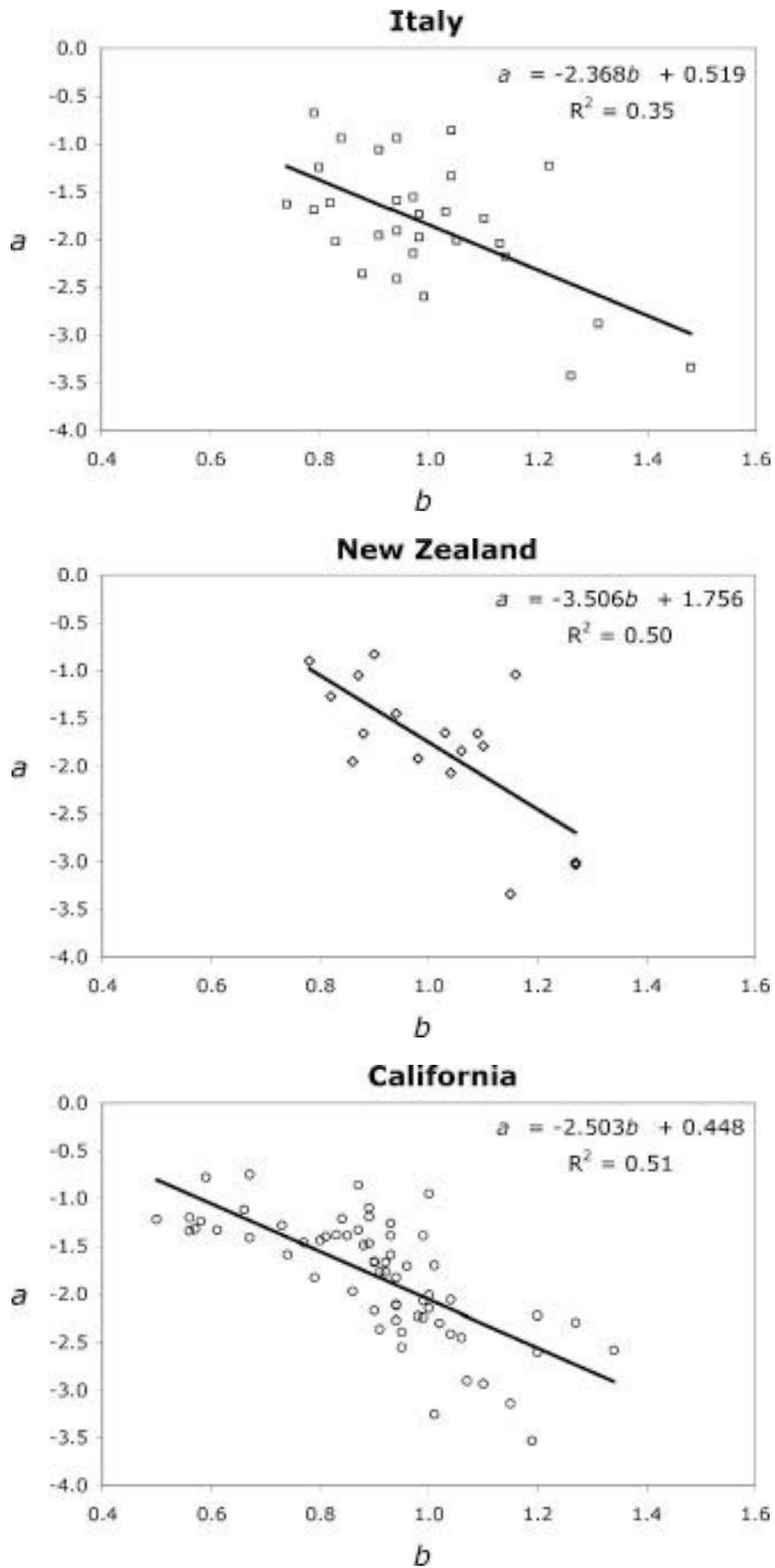

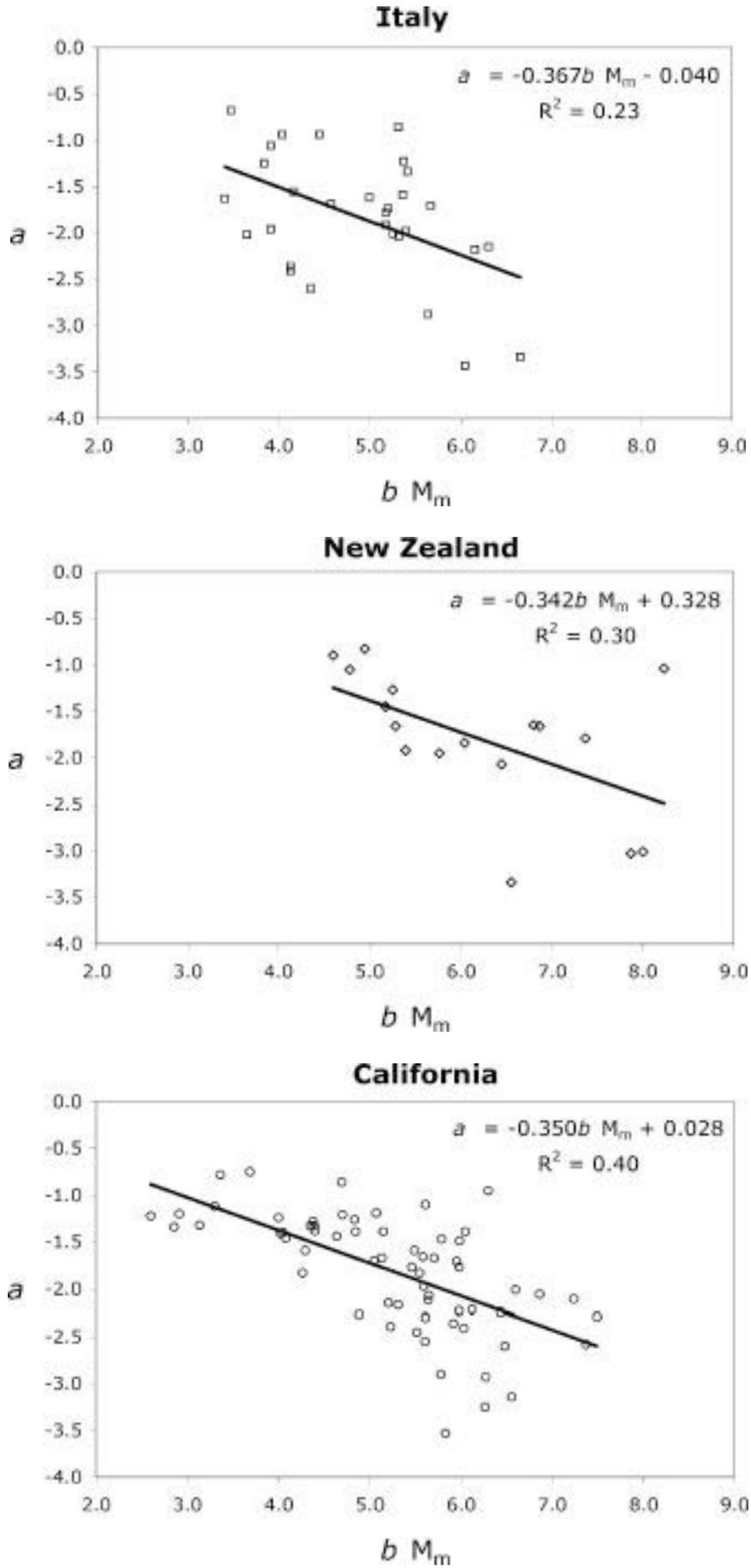

**Figure 3**

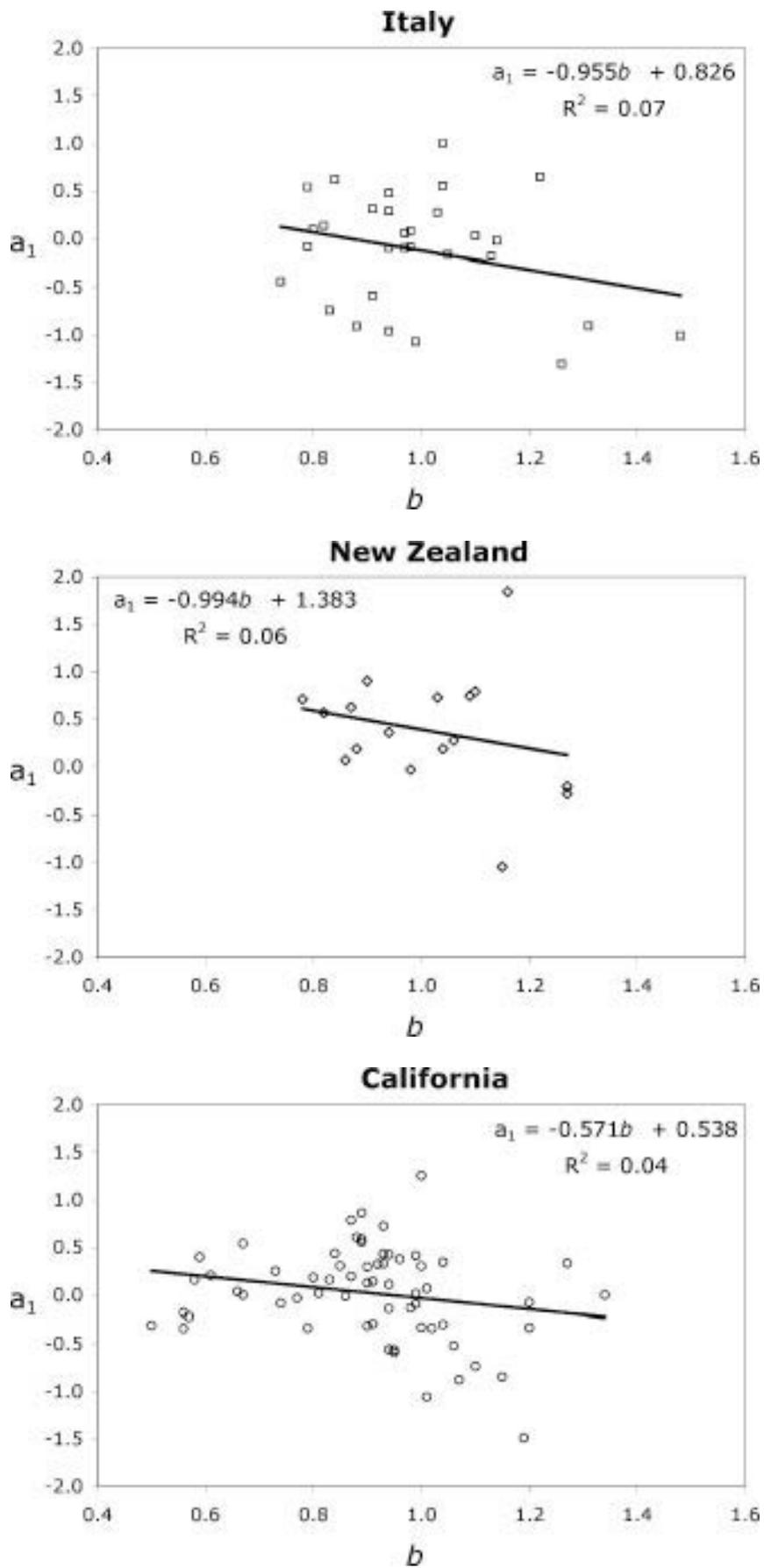

**Figure 4**

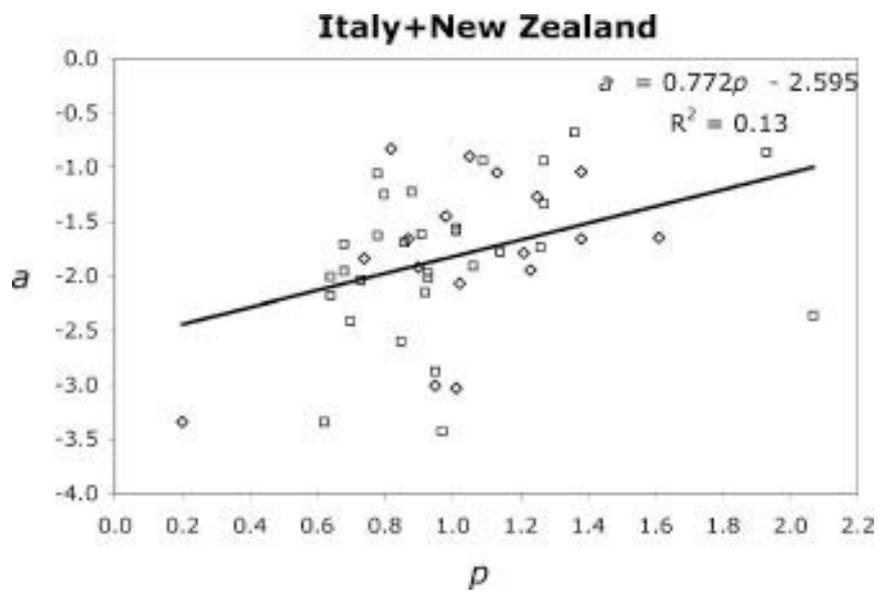

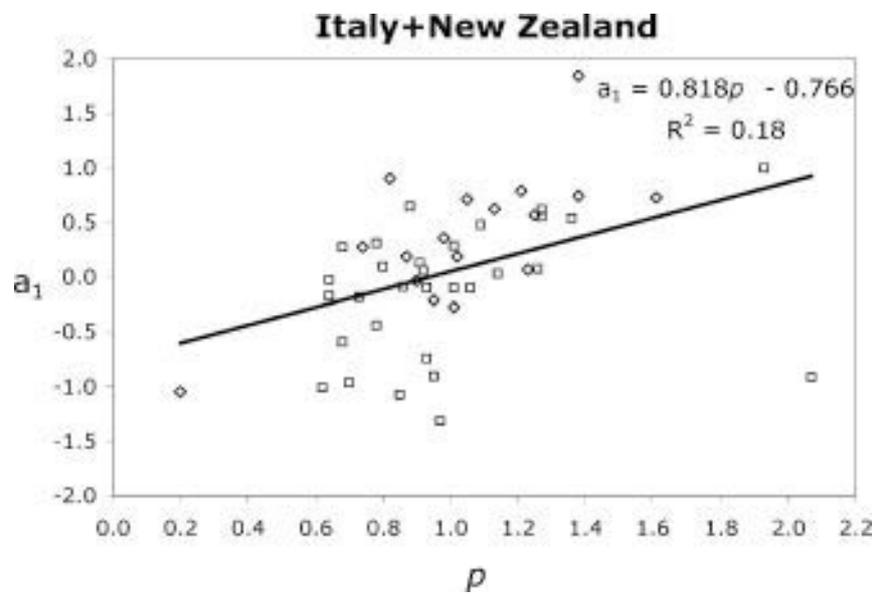

**Figure 5**

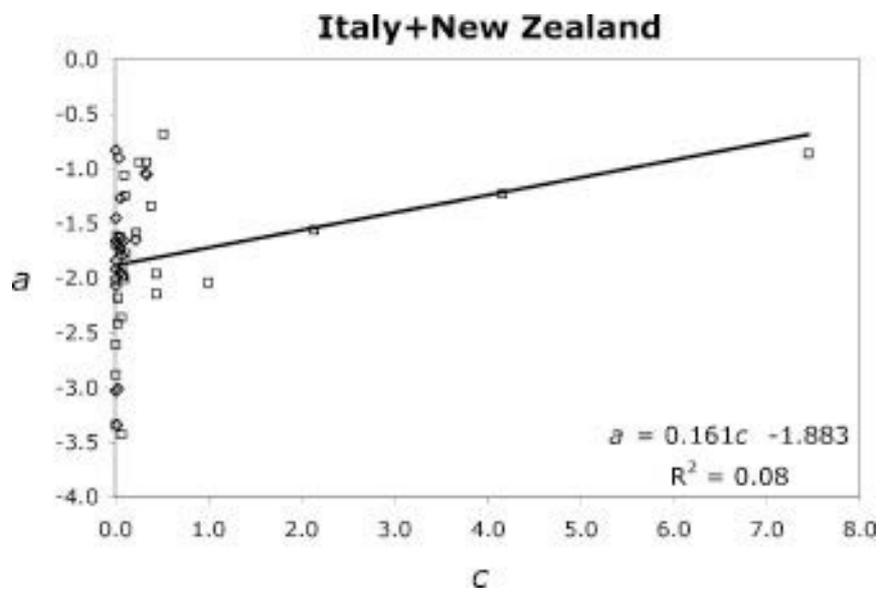

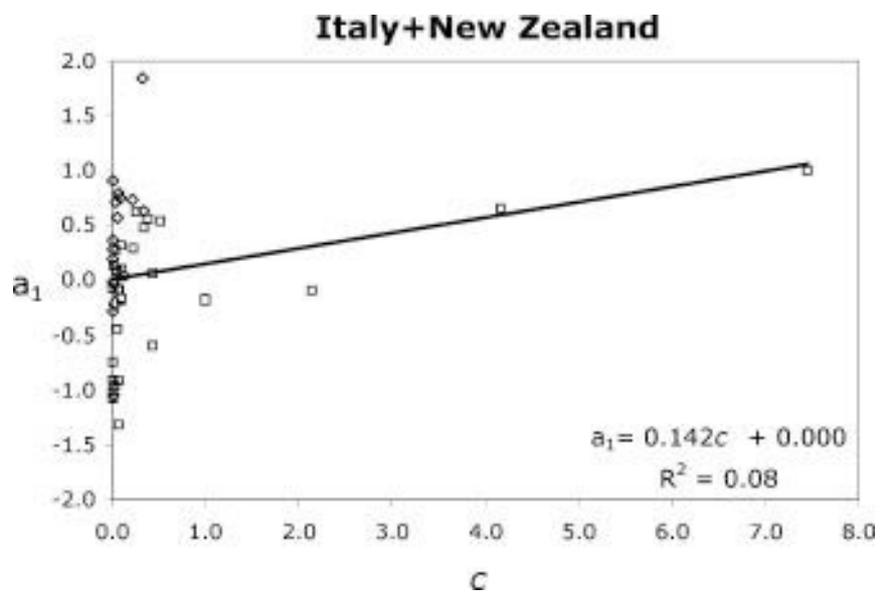

**Figure 6**

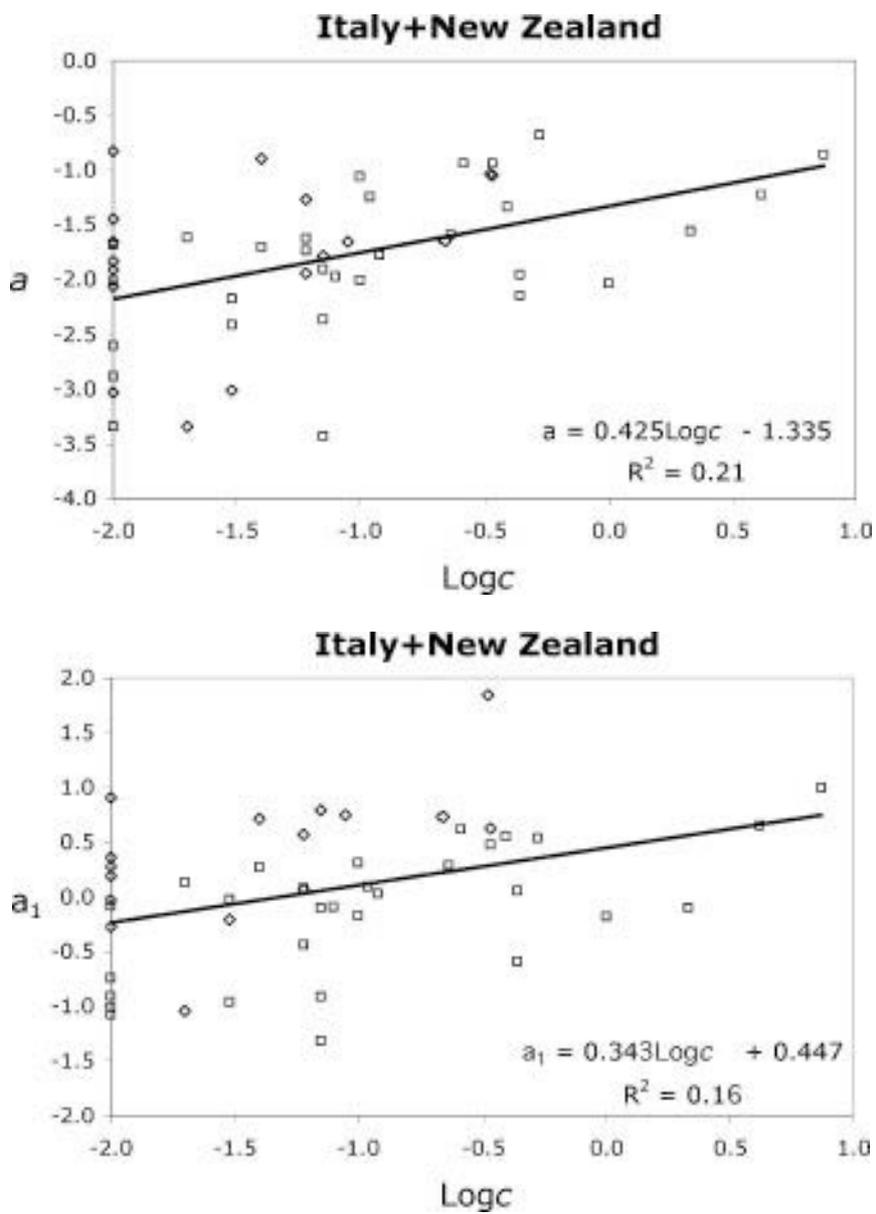

**Figure 7**

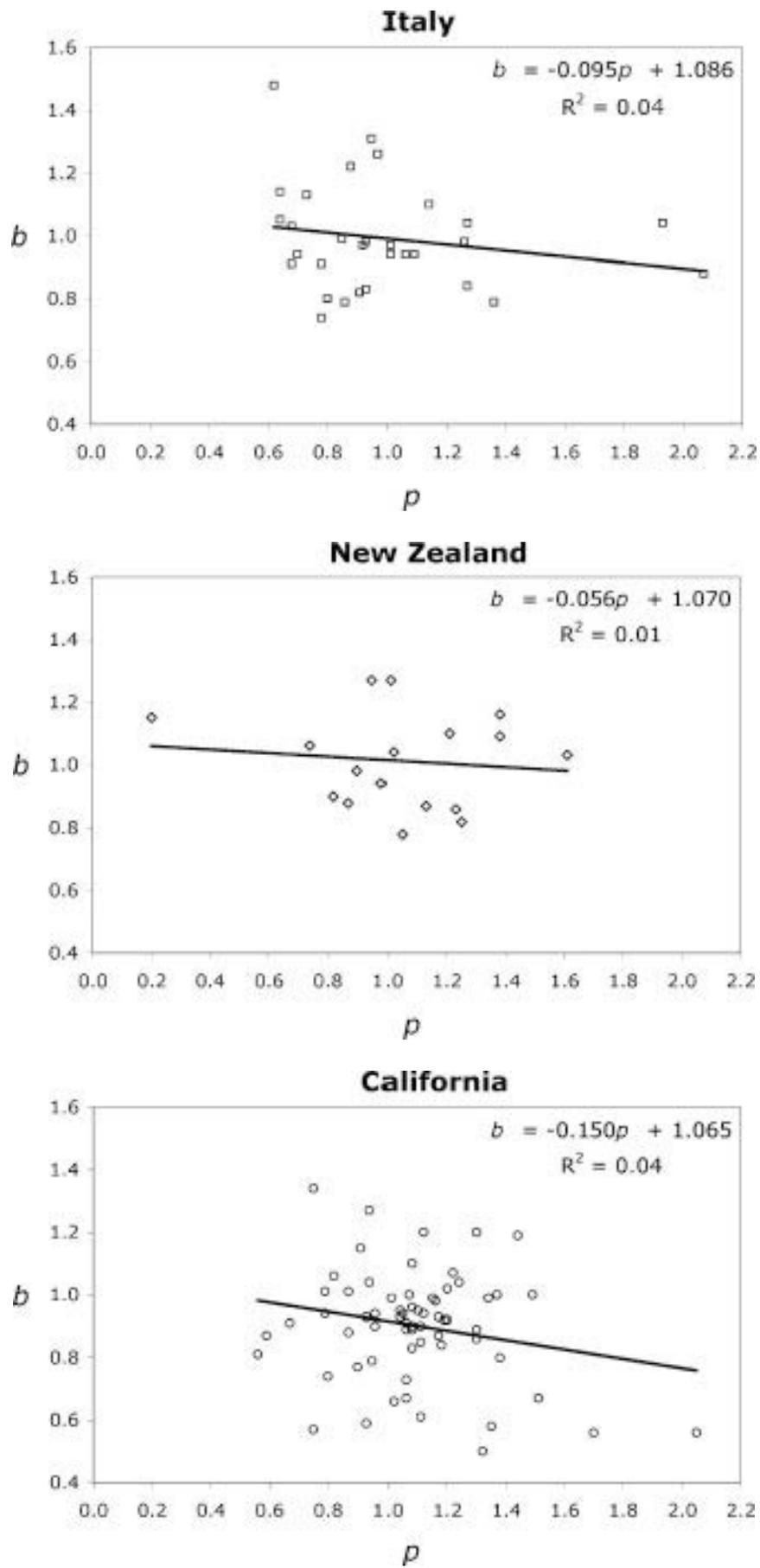

**Figure 8**

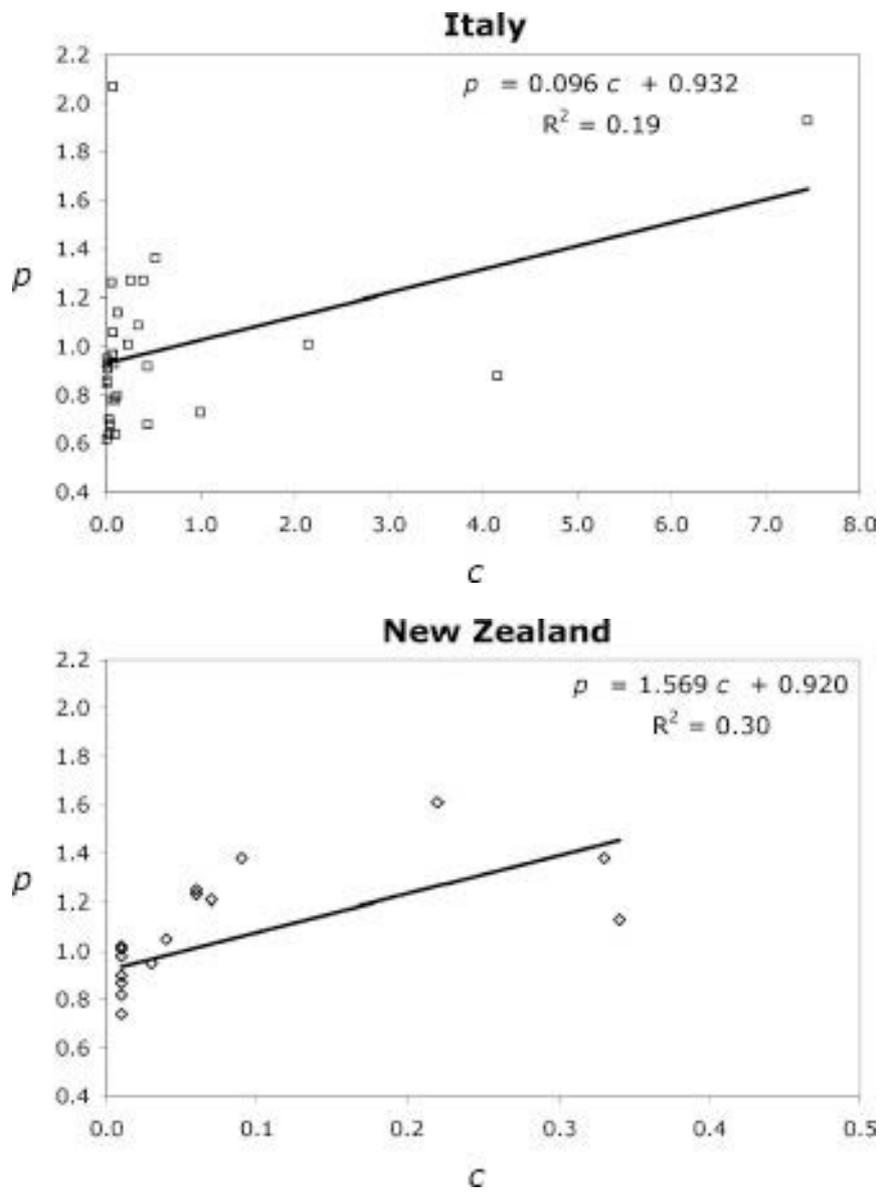

**Figure 9**

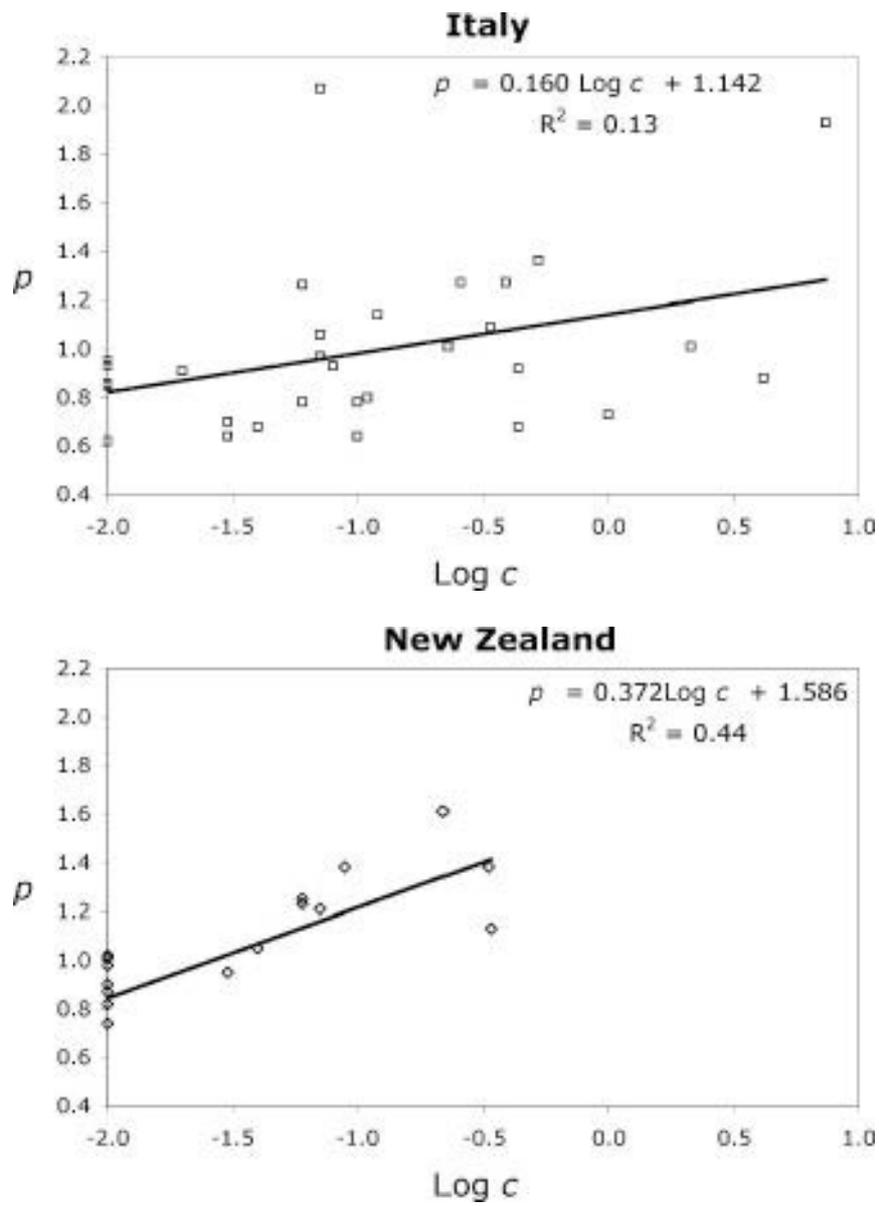

**Figure 10**

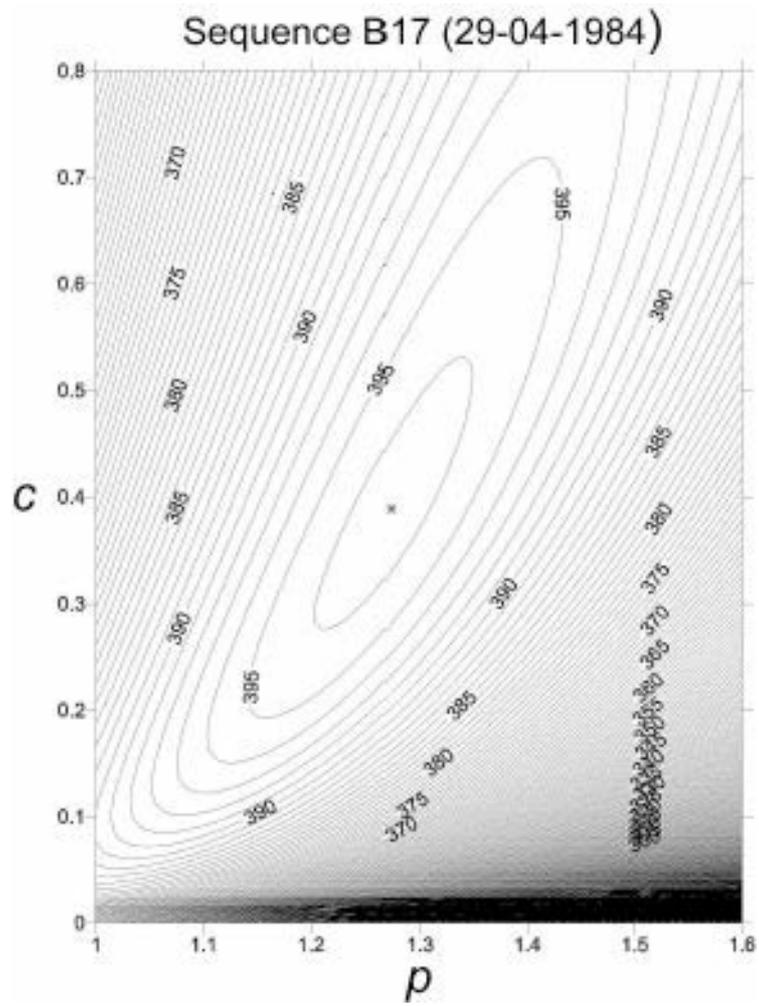

# Table captions

Table 1 – Correlation analysis between productivity parameters $a$, $a_1$ and $a_2$ with $b$, $bM_m$ and $\alpha M_m$. "Coeff" and "Coeff std" indicate the regression coefficient and its standard deviation, "Intcep" and "Intcep std" the regression intercept term with its standard deviation, "r" the correlation coefficient, "t" the value of Student-t statistics for the null hypothesis of absence of correlation, "d.f." and "s.l." the degrees of freedom and the significance level of t-test, column "corr" reports the sign (positive or negative) of correlation, when significant, or blank otherwise. "New Zealand (Ml)" in the dataset column, indicates that $M_m$, $M_{min}$ and $b$ values for New Zealand are converted from Mw to Ml to conform other regions.

Table 2 – Correlation analysis between $a$ and $bM_m$ using bootstrap resampling technique. For each dataset, the first row reports the results of standard linear regression, the second one the average values computed over the set of 5000 resamplings and the third one the median values. Regression coefficient uncertainties are computed in the second row as standard deviations of the distribution of values obtained from resampling, while in the third and fourth ones as the difference between the median and the 16[th] and 84[th] percentiles of the same distribution of values.

Table 3 – Averages and standard deviations (std) of magnitude independent productivity $a$ and its transformations $a_0$, $a_1$ and $a_2$ (see text) for the three region separately and for the joint dataset of all thee region together. Columns "$a$-$a_0$ std", "$a$-$a_1$ std", "$a_1$-$a_2$ std" and "$a$-$a_2$ std" report the standard deviation variation between corresponding parameters in percentage. "New Zealand (Ml)" and "Joint (Ml)" in the dataset column indicate that $M_m$, $M_{min}$ and $b$ values for New Zealand are converted from Mw to Ml to conform other regions.

Table 4 – Correlation analysis of productivity parameters $a$, $a_0$, $a_1$ and $a_2$ with $p$, $c$ and Log $c$. The explanation of different columns is the same as of Table 1.

Table 5 – Correlation analysis between $b$, $p$, $c$ and Log $c$. The explanation of different columns is the same as of Table 1.

Table 6 – List of Italian sequences used for the validation of the new rate formula. Lat, Lon and $M_m$ are the coordinates and the magnitude of the mainshock, $M_{min}$ and N the minimum magnitude and the total number of aftershocks, S and T the starting and ending time of the modeled sequence (in days), $LR_{AV}$, $LR_{WA}$, $LR_{MD}$ the log-likelihood ratios between the original Reasenberg and Jones (1989) model and the new model defined by eq. (13) and (14) when the values of model parameters used are computed as arithmetic average, weighted average and median respectively.

Table 7 – Values of average parameters used for the validation of the new rate formula.

Table 8 – Results of the application the epidemic type model of eq. (17) to seven Italian sequences. "Seq." indicates the sequence code assigned by Lolli and Gasperini (2003), $M_m$ the mainshock magnitude, $M_{min}$ the magnitude minimum threshold, $N$ the total number of shock above $M_{min}$, $p_{MOM}$, $c_{MOM}$, $b_{MOM}$ and $a_{MOM}$ the estimates of the Reasenberg and Jones (1989) model parameters made by Lolli and Gasperini (2003), $p_{ETAS}$, $c_{ETAS}$, $b_{ETAS}$, $\alpha$ and $a_1$ the estimates of the ETAS model parameters (eq. 17), std $b$ and std $\alpha$ the standard deviation of ETAS parameters $b_{ETAS}$, $\alpha$ respectively.

**Table 1**

| Params | Dataset | Coeff | Coeff std | Intcep | Intcep std | r | t | d.f. | s.l. | corr |
|---|---|---|---|---|---|---|---|---|---|---|
| $a\ b$ | Italy+New Zealand+California | -2.387 | 0.267 | 0.439 | 0.256 | -0.65 | 8.94 | 109 | <0.01 | negative |
| $a\ b$ | Italy+New Zealand | -2.698 | 0.499 | 0.880 | 0.505 | -0.63 | 5.41 | 45 | <0.01 | negative |
| $a\ b$ | Italy | -2.368 | 0.609 | 0.519 | 0.612 | -0.59 | 3.89 | 28 | <0.01 | negative |
| $a\ b$ | New Zealand | -3.506 | 0.898 | 1.756 | 0.918 | -0.71 | 3.90 | 15 | <0.01 | negative |
| $a\ b$ | California | -2.503 | 0.309 | 0.448 | 0.284 | -0.72 | 8.09 | 62 | <0.01 | negative |
| $a\ bM_m$ | Italy+New Zealand+California | -0.301 | 0.046 | -0.222 | 0.251 | -0.53 | 6.47 | 109 | <0.01 | negative |
| $a\ bM_m$ | Italy+New Zeal.+Calif. M≥6.0 | -0.237 | 0.082 | -0.386 | 0.502 | -0.45 | 2.88 | 33 | <0.01 | negative |
| $a\ bM_m$ | Italy+New Zealand | -0.241 | 0.080 | -0.522 | 0.438 | -0.41 | 3.02 | 45 | <0.01 | negative |
| $a\ bM_m$ | Italy | -0.367 | 0.126 | -0.040 | 0.627 | -0.48 | 2.90 | 28 | <0.01 | negative |
| $a\ bM_m$ | New Zealand | -0.342 | 0.136 | 0.328 | 0.857 | -0.54 | 2.52 | 15 | 0.02 | negative |
| $a\ bM_m$ | California | -0.351 | 0.055 | 0.028 | 0.295 | -0.63 | 6.37 | 62 | <0.01 | negative |
| $a\ bM_m$ | Italy+NewZealand (Ml)+Calif. | -0.346 | 0.049 | -0.034 | 0.257 | -0.56 | 7.05 | 109 | <0.01 | negative |
| $a\ bM_m$ | Italy+NewZeal. (Ml)+Calif. M≥6.0 | -0.305 | 0.092 | -0.054 | 0.530 | -0.49 | 3.31 | 34 | <0.01 | negative |
| $a\ bM_m$ | Italy+New Zealand (Ml) | -0.348 | 0.095 | -0.072 | 0.483 | -0.48 | 3.67 | 45 | <0.01 | negative |
| $a\ bM_m$ | New Zealand (Ml) | -0.362 | 0.155 | 0.115 | 0.833 | -0.52 | 2.33 | 15 | 0.03 | negative |
| $a\ bM_{min}$ | Italy+New Zealand+California | -0.236 | 0.079 | -1.246 | 0.199 | -0.27 | 2.98 | 109 | <0.01 | negative |
| $a\ bM_{min}$ | Italy+New Zealand | -0.080 | 0.127 | -1.637 | 0.301 | -0.09 | 0.63 | 45 | 0.53 | |
| $a\ bM_{min}$ | Italy | -0.149 | 0.263 | -1.553 | 0.504 | -0.11 | 0.57 | 28 | 0.58 | |
| $a\ bM_{min}$ | New Zealand | -0.151 | 0.217 | -1.359 | 0.650 | -0.18 | 0.69 | 15 | 0.50 | |
| $a\ bM_{min}$ | California | -0.414 | 0.101 | -0.762 | 0.266 | -0.46 | 4.08 | 62 | <0.01 | negative |
| $a_1\ b$ | Italy+New Zealand+California | -0.605 | 0.295 | 0.609 | 0.283 | -0.19 | 2.05 | 109 | 0.04 | negative |
| $a_1\ b$ | Italy+New Zealand | -0.873 | 0.582 | 0.932 | 0.589 | -0.22 | 1.50 | 45 | 0.14 | |
| $a_1\ b$ | Italy | -0.955 | 0.638 | 0.826 | 0.641 | -0.27 | 1.50 | 28 | 0.15 | |
| $a_1\ b$ | New Zealand | -0.994 | 1.038 | 1.383 | 1.061 | -0.24 | 0.96 | 15 | 0.35 | |
| $a_1\ b$ | California | -0.571 | 0.337 | 0.538 | 0.310 | -0.21 | 1.69 | 62 | 0.10 | |
| $a_1\ \alpha M_m$ | Italy+New Zealand+California | 0.076 | 0.071 | -0.222 | 0.251 | 0.10 | 1.06 | 109 | 0.29 | |
| $a_1\ \alpha M_m$ | Italy+New Zealand | 0.167 | 0.123 | -0.522 | 0.438 | 0.20 | 1.36 | 45 | 0.18 | |
| $a_1\ \alpha M_m$ | Italy | -0.025 | 0.195 | -0.040 | 0.627 | -0.02 | 0.13 | 28 | 0.90 | |
| $a_1\ \alpha M_m$ | New Zealand | 0.012 | 0.209 | 0.328 | 0.857 | 0.02 | 0.06 | 15 | 0.95 | |
| $a_1\ \alpha M_m$ | California | -0.001 | 0.085 | 0.028 | 0.295 | 0.00 | 0.02 | 62 | 0.99 | |
| $a_2\ b$ | Italy+New Zealand+California | -0.509 | 0.274 | 1.552 | 0.262 | -0.18 | 1.86 | 109 | 0.07 | |
| $a_2\ b$ | Italy+New Zealand | -0.511 | 0.485 | 1.616 | 0.491 | -0.16 | 1.05 | 45 | 0.30 | |
| $a_2\ b$ | Italy | -0.670 | 0.543 | 1.582 | 0.546 | -0.23 | 1.23 | 28 | 0.23 | |
| $a_2\ b$ | New Zealand | -0.464 | 0.709 | 1.907 | 0.724 | -0.17 | 0.65 | 15 | 0.52 | |
| $a_2\ b$ | California | -0.653 | 0.350 | 1.636 | 0.321 | -0.23 | 1.87 | 62 | 0.07 | |
| $a_2\ \alpha M_m$ | Italy+New Zealand+California | 0.053 | 0.066 | 0.891 | 0.233 | 0.08 | 0.79 | 109 | 0.43 | |
| $a_2\ \alpha M_m$ | Italy+New Zealand | 0.168 | 0.100 | 0.521 | 0.357 | 0.24 | 1.68 | 45 | 0.10 | |
| $a_2\ \alpha M_m$ | Italy | 0.006 | 0.164 | 0.899 | 0.528 | 0.01 | 0.04 | 28 | 0.97 | |
| $a_2\ \alpha M_m$ | New Zealand | -0.039 | 0.140 | 1.593 | 0.575 | -0.07 | 0.27 | 15 | 0.79 | |
| $a_2\ \alpha M_m$ | California | -0.046 | 0.088 | 1.204 | 0.306 | -0.07 | 0.52 | 62 | 0.60 | |

**Table 2**

| Method | Dataset | Coeff | Coeff uncert. | Intcep | Intcep uncert. | r | t | d.f. | s.l. |
|---|---|---|---|---|---|---|---|---|---|
| Lin reg | Italy+New Zealand+California | -0.301 | 0.046 | -0.222 | 0.251 | -0.53 | 6.475 | 109 | <0.01 |
| Boot mean | Italy+New Zealand+California | -0.304 | 0.052 | -0.208 | 0.264 | -0.53 | 6.614 | 109 | <0.01 |
| Boot median | Italy+New Zealand+California | -0.303 | 0.052 | -0.205 | 0.264 | -0.53 | 6.578 | 109 | <0.01 |
| | | | 0.052 | | 0.258 | | | | |
| Lin reg | Italy+New Zealand | -0.241 | 0.080 | -0.522 | 0.438 | -0.41 | 3.025 | 45 | <0.01 |
| Boot mean | Italy+New Zealand | -0.247 | 0.099 | -0.492 | 0.510 | -0.41 | 2.155 | 45 | 0.02 |
| Boot median | Italy+New Zealand | -0.244 | 0.103 | -0.505 | 0.497 | -0.42 | 3.077 | 45 | <0.01 |
| | | | 0.096 | | 0.526 | | | | |
| Lin reg | Italy | -0.367 | 0.126 | -0.040 | 0.627 | -0.48 | 2.899 | 28 | <0.01 |
| Boot mean | Italy | -0.355 | 0.141 | -0.091 | 0.683 | -0.46 | 2.933 | 28 | <0.01 |
| Boot median | Italy | -0.362 | 0.135 | -0.068 | 0.690 | -0.48 | 2.922 | 28 | <0.01 |
| | | | 0.147 | | 0.657 | | | | |
| Lin reg | New Zealand | -0.342 | 0.136 | 0.328 | 0.857 | -0.54 | 2.515 | 15 | 0.01 |
| Boot mean | New Zealand | -0.352 | 0.167 | 0.370 | 0.946 | -0.55 | 3.092 | 15 | <0.01 |
| Boot median | New Zealand | -0.356 | 0.168 | 0.412 | 1.020 | -0.57 | 2.711 | 15 | <0.01 |
| | | | 0.178 | | 0.938 | | | | |
| Lin reg | California | -0.351 | 0.055 | 0.028 | 0.295 | -0.63 | 6.371 | 62 | <0.01 |
| Boot mean | California | -0.354 | 0.048 | 0.045 | 0.233 | -0.63 | 6.494 | 62 | <0.01 |
| Boot median | California | -0.352 | 0.051 | 0.028 | 0.209 | -0.63 | 6.440 | 62 | <0.01 |
| | | | 0.045 | | 0.247 | | | | |

**Table 3**

| Dataset | Nseq | $a$ | $a$ std | $a_0$ | $a_0$ std | $a-a_0$ std | $a_1$ | $a_1$ std | $a-a_1$ std | $a_2$ | $a_2$ std | $a_1-a_2$ std | $a-a_2$ std |
|---|---|---|---|---|---|---|---|---|---|---|---|---|---|
| Italy | 30 | -1.830 | 0.146 | -0.703 | 0.127 | -13.0% | -0.121 | 0.125 | -14.1% | 0.917 | 0.094 | -24.6% | -35.3% |
| New Zealand | 17 | -1.792 | 0.186 | -0.665 | 0.156 | -16.1% | 0.378 | 0.156 | -16.1% | 1.438 | 0.105 | -32.8% | -43.6% |
| New Zealand (MI) | 17 | -1.792 | 0.186 | -0.665 | 0.156 | -16.1% | -0.028 | 0.157 | -15.6% | 1.032 | 0.108 | -31.2% | -42.0% |
| California | 64 | -1.809 | 0.078 | -0.737 | 0.079 | +0.8% | 0.023 | 0.061 | -22.3% | 1.048 | 0.063 | +4.2% | -19.0% |
| Joint | 111 | -1.812 | 0.062 | -0.717 | 0.061 | -1.5% | 0.039 | 0.053 | -14.6% | 1.072 | 0.049 | -7.5% | -20.9% |
| Joint (MI) | 111 | -1.812 | 0.062 | -0.717 | 0.061 | -1.5% | -0.024 | 0.051 | -17.5% | 1.010 | 0.047 | -8.5% | -24.5% |

**Table 4**

| Params | Dataset | Coeff | Coeff std | Intcep | Intcep std | r | t | d.f. | s.l. | corr |
|---|---|---|---|---|---|---|---|---|---|---|
| $a\ p$ | Italy+New Zealand+California | 0.509 | 0.210 | -2.351 | 0.230 | 0.23 | 2.43 | 109 | 0.02 | positive |
| $a\ p$ | Italy+New Zealand | 0.772 | 0.291 | -2.595 | 0.309 | 0.37 | 2.65 | 45 | 0.01 | positive |
| $a\ p$ | Italy | 0.610 | 0.355 | -2.433 | 0.371 | 0.31 | 1.72 | 28 | 0.10 | |
| $a\ p$ | New Zealand | 1.120 | 0.537 | -2.959 | 0.584 | 0.47 | 2.08 | 15 | >0.05 | |
| $a\ p$ | California | 0.199 | 0.312 | -2.026 | 0.350 | 0.08 | 0.64 | 62 | 0.53 | |
| $a\ c$ | Italy+New Zealand | 0.161 | 0.079 | -1.883 | 0.103 | 0.29 | 2.03 | 45 | <0.05 | positive |
| $a\ c$ | Italy | 0.164 | 0.077 | -1.930 | 0.125 | 0.37 | 2.13 | 28 | 0.04 | positive |
| $a\ c$ | New Zealand | 2.756 | 1.595 | -2.007 | 0.211 | 0.41 | 1.73 | 15 | 0.10 | |
| $a\ \text{Log}c$ | Italy+New Zealand | 0.425 | 0.123 | -1.335 | 0.166 | 0.46 | 3.45 | 45 | <0.01 | positive |
| $a\ \text{Log}c$ | Italy | 0.497 | 0.135 | -1.361 | 0.163 | 0.57 | 3.69 | 28 | <0.01 | positive |
| $a\ \text{Log}c$ | New Zealand | 0.467 | 0.319 | -1.109 | 0.497 | 0.35 | 1.46 | 15 | 0.16 | |
| $a_1\ p$ | Italy+New Zealand+California | 0.420 | 0.179 | -0.405 | 0.197 | 0.22 | 2.34 | 109 | 0.02 | positive |
| $a_1\ p$ | Italy+New Zealand | 0.818 | 0.265 | -0.766 | 0.281 | 0.42 | 3.09 | 45 | <0.01 | positive |
| $a_1\ p$ | Italy | 0.479 | 0.315 | -0.595 | 0.329 | 0.28 | 1.52 | 28 | 0.14 | |
| $a_1\ p$ | New Zealand | 1.379 | 0.368 | -1.061 | 0.399 | 0.70 | 3.75 | 15 | <0.01 | positive |
| $a_1\ p$ | California | -0.042 | 0.244 | 0.070 | 0.273 | -0.02 | 0.17 | 62 | 0.86 | |
| $a_1\ c$ | Italy+New Zealand | 0.142 | 0.074 | 0.000 | 0.096 | 0.27 | 1.92 | 45 | 0.06 | |
| $a_1\ c$ | Italy | 0.181 | 0.064 | -0.232 | 0.105 | 0.47 | 2.82 | 28 | <0.01 | positive |
| $a_1\ c$ | New Zealand | 3.405 | 1.172 | 0.111 | 0.155 | 0.60 | 2.91 | 15 | 0.01 | positive |
| $a_1\ \text{Log}c$ | Italy+New Zealand | 0.343 | 0.118 | 0.447 | 0.160 | 0.40 | 2.90 | 45 | <0.01 | positive |
| $a_1\ \text{Log}c$ | Italy | 0.480 | 0.112 | 0.332 | 0.136 | 0.63 | 4.28 | 28 | <0.01 | positive |
| $a_1\ \text{Log}c$ | New Zealand | 0.625 | 0.236 | 1.293 | 0.368 | 0.57 | 2.65 | 15 | 0.02 | positive |
| $a_0\ p$ | Italy+New Zealand+California | -0.425 | 0.208 | -0.267 | 0.228 | -0.19 | 2.04 | 109 | 0.04 | negative |
| $a_0\ p$ | Italy+New Zealand | -0.472 | 0.263 | -0.214 | 0.279 | -0.26 | 1.79 | 45 | 0.08 | |
| $a_0\ p$ | Italy | -0.493 | 0.318 | -0.215 | 0.332 | -0.28 | 1.55 | 28 | 0.13 | |
| $a_0\ p$ | New Zealand | -0.448 | 0.499 | -0.198 | 0.542 | -0.23 | 0.90 | 15 | 0.38 | |
| $a_0\ p$ | California | -0.360 | 0.336 | -0.343 | 0.376 | -0.14 | 1.07 | 62 | 0.29 | |
| $a_0\ c$ | Italy+New Zealand | -0.055 | 0.071 | -0.667 | 0.093 | -0.11 | 0.77 | 45 | 0.45 | |
| $a_0\ c$ | Italy | -0.056 | 0.073 | -0.669 | 0.119 | -0.14 | 0.76 | 28 | 0.45 | |
| $a_0\ c$ | New Zealand | 0.473 | 1.461 | -0.702 | 0.193 | 0.08 | 0.32 | 15 | 0.75 | |
| $a_0\ \text{Log}c$ | Italy+New Zealand | 0.083 | 0.120 | -0.595 | 0.162 | 0.10 | 0.70 | 45 | 0.49 | |
| $a_0\ \text{Log}c$ | Italy | 0.138 | 0.143 | -0.573 | 0.174 | 0.18 | 0.96 | 28 | 0.34 | |
| $a_0\ \text{Log}c$ | New Zealand | -0.014 | 0.286 | -0.686 | 0.446 | -0.01 | 0.05 | 15 | 0.96 | |
| $a_2\ p$ | Italy+New Zealand+California | -0.327 | 0.167 | 1.418 | 0.183 | -0.18 | 1.95 | 109 | >0.05 | |
| $a_2\ p$ | Italy+New Zealand | -0.231 | 0.238 | 1.338 | 0.252 | -0.14 | 0.97 | 45 | 0.34 | |
| $a_2\ p$ | Italy | -0.471 | 0.261 | 1.384 | 0.273 | -0.32 | 1.80 | 28 | 0.08 | |
| $a_2\ p$ | New Zealand | 0.091 | 0.343 | 1.343 | 0.373 | 0.07 | 0.27 | 15 | 0.79 | |
| $a_2\ p$ | California | -0.430 | 0.248 | 1.518 | 0.278 | -0.22 | 1.74 | 62 | 0.09 | |
| $a_2\ c$ | Italy+New Zealand | -0.067 | 0.063 | 1.134 | 0.082 | -0.16 | 1.08 | 45 | 0.29 | |
| $a_2\ c$ | Italy | -0.031 | 0.061 | 0.936 | 0.099 | -0.09 | 0.50 | 28 | 0.62 | |
| $a_2\ c$ | New Zealand | 1.408 | 0.915 | 1.328 | 0.121 | 0.37 | 1.54 | 15 | 0.14 | |
| $a_2\ \text{Log}c$ | Italy+New Zealand | 0.021 | 0.106 | 1.130 | 0.143 | 0.03 | 0.20 | 45 | 0.84 | |
| $a_2\ \text{Log}c$ | Italy | 0.138 | 0.118 | 1.048 | 0.144 | 0.22 | 1.17 | 28 | 0.25 | |
| $a_2\ \text{Log}c$ | New Zealand | 0.211 | 0.184 | 1.746 | 0.287 | 0.28 | 1.14 | 15 | 0.27 | |

**Table 5**

| Params | Dataset | Coeff | Coeff std | Intcep | Intcep std | r | t | d.f. | s.l. | corr |
|---|---|---|---|---|---|---|---|---|---|---|
| $b\ p$ | Italy+New Zealand+California | -0.133 | 0.057 | 1.083 | 0.063 | -0.22 | 2.32 | 109 | 0.02 | negative |
| $b\ p$ | Italy+New Zealand | -0.080 | 0.072 | 1.080 | 0.076 | -0.16 | 1.11 | 45 | 0.27 | |
| $b\ p$ | Italy | -0.095 | 0.092 | 1.086 | 0.096 | -0.19 | 1.04 | 28 | 0.31 | |
| $b\ p$ | New Zealand | -0.056 | 0.123 | 1.070 | 0.133 | -0.12 | 0.46 | 15 | 0.66 | |
| $b\ p$ | California | -0.150 | 0.088 | 1.065 | 0.098 | -0.21 | 1.71 | 62 | 0.09 | |
| $b\ c$ | Italy+New Zealand | 0.015 | 0.019 | 0.993 | 0.025 | 0.12 | 0.78 | 45 | 0.44 | |
| $b\ c$ | Italy | 0.017 | 0.021 | 0.981 | 0.033 | 0.16 | 0.84 | 28 | 0.41 | |
| $b\ c$ | New Zealand | -0.008 | 0.354 | 1.012 | 0.047 | -0.01 | 0.02 | 15 | 0.98 | |
| $b\ \text{Log}c$ | Italy+New Zealand | -0.009 | 0.032 | 0.989 | 0.043 | -0.04 | 0.29 | 45 | 0.78 | |
| $b\ \text{Log}c$ | Italy | -0.004 | 0.041 | 0.989 | 0.050 | -0.02 | 0.09 | 28 | 0.93 | |
| $b\ \text{Log}c$ | New Zealand | -0.012 | 0.069 | 0.995 | 0.108 | -0.04 | 0.17 | 15 | 0.87 | |
| $p\ c$ | Italy+New Zealand | 0.091 | 0.037 | 0.971 | 0.048 | 0.35 | 2.48 | 45 | 0.02 | positive |
| $p\ c$ | Italy | 0.096 | 0.038 | 0.932 | 0.062 | 0.43 | 2.53 | 28 | 0.02 | positive |
| $p\ c$ | New Zealand | 1.569 | 0.618 | 0.920 | 0.082 | 0.55 | 2.54 | 15 | 0.02 | positive |
| $p\ \text{Log}c$ | Italy+New Zealand | 0.172 | 0.061 | 1.204 | 0.082 | 0.39 | 2.83 | 45 | <0.01 | positive |
| $p\ \text{Log}c$ | Italy | 0.160 | 0.077 | 1.142 | 0.094 | 0.36 | 2.07 | 28 | <0.05 | positive |
| $p\ \text{Log}c$ | New Zealand | 0.371 | 0.108 | 1.586 | 0.168 | 0.67 | 3.45 | 15 | <0.01 | positive |

**Table 6**

| Seq | dd/mm/yy | Lat | Lon | $M_m$ | $M_{min}$ | N | S | T | $LR_{AV}$ | $LR_{WA}$ | $LR_{MD}$ |
|---|---|---|---|---|---|---|---|---|---|---|---|
| 1 | 12/05/97 | 42 46 | 12 37 | 4.5 | 2.0 | 104 | 0.01 | 132.8 | **-12.3** | **-13.1** | **-19.8** |
| 2 | 12/04/98 | 46 24 | 13 48 | 5.5 | 2.0 | 89 | 0.01 | 139.4 | **-89.8** | **-114.9** | **-63.9** |
| 3 | 20/06/98 | 38 23 | 12 50 | 4.3 | 2.0 | 99 | 0.01 | 55.2 | **-26.5** | **-27.7** | **-34.8** |
| 4 | 09/09/98 | 40 15 | 16 13 | 5.6 | 2.0 | 146 | 0.01 | 226.9 | **-130.8** | **-175.6** | **-94.1** |
| 5 | 14/02/99 | 38 14 | 15 00 | 4.6 | 1.6 | 24 | 0.03 | 30.8 | 5.5 | 16.1 | 12.3 |
| 6 | 07/07/99 | 44 15 | 10 49 | 4.9 | 2.0 | 38 | 0.02 | 177.8 | **-5.9** | **-3.6** | 0.0 |
| 7 | 29/12/99 | 46 24 | 10 23 | 4.7 | 1.5 | 143 | 0.01 | 81.6 | **-0.1** | 3.5 | **-0.7** |
| 8 | 10/05/00 | 44 13 | 11 58 | 4.7 | 1.7 | 240 | 0.01 | 177.8 | **-3.2** | **-9.3** | **-19.8** |
| 9 | 22/06/00 | 43 21 | 12 25 | 4.5 | 1.5 | 55 | 0.01 | 169.4 | 9.0 | 20.4 | 15.3 |
| 10 | 21/08/00 | 44 51 | 08 21 | 5.2 | 2.0 | 20 | 0.09 | 250.1 | **-40.6** | **-40.2** | **-24.7** |
| 11 | 18/06/02 | 44 21 | 10 43 | 4.4 | 2.5 | 26 | 0.01 | 427.3 | **-3.7** | **-3.0** | **-4.9** |
| 12 | 06/09/02 | 38 25 | 13 43 | 5.6 | 3.0 | 74 | 0.01 | 132.8 | 29.4 | 16.9 | 21.9 |
| 13 | 27/10/02 | 37 43 | 15 07 | 4.8 | 3.0 | 33 | 0.01 | 139.4 | 1.5 | **-0.7** | **-2.0** |
| 14 | 31/10/02 | 41 43 | 14 54 | 5.4 | 3.5 | 30 | 0.01 | 369.2 | 12.1 | 8.5 | 8.3 |
| 15 | 26/01/03 | 43 53 | 11 55 | 4.3 | 2.5 | 34 | 0.01 | 120.4 | **-9.0** | **-9.2** | **-11.6** |
| 16 | 29/03/03 | 43 07 | 15 27 | 5.4 | 3.0 | 79 | 0.01 | 19.9 | 34.2 | 24.6 | 24.1 |
| 17 | 29/04/03 | 43 06 | 15 22 | 4.4 | 2.2 | 56 | 0.01 | 153.6 | **-10.3** | **-11.1** | **-14.5** |
| Cumulative | | | | | | | | | -240.4 | -318.3 | -208.9 |

**Table 7**

|  | p | Log c | c | b | a | $a_1$ | $\alpha$ |
|---|---|---|---|---|---|---|---|
| arithmetic average | 0.989 | -0.936 | 0.116 | 0.994 | -1.828 | -0.182 | 0.646 |
| weighted average | 0.932 | -1.531 | 0.029 | 0.963 | -1.656 | -0.026 | 0.626 |
| median | 0.930 | -1.048 | 0.089 | 0.955 | -1.735 | -0.097 | 0.621 |

**Table 8**

| Seq. | Date | $M_m$ | $M_{min}$ | N | $p_{MOM}$ | $c_{MOM}$ | $a_{MOM}$ | $b_{MOM}$ | $p_{ETAS}$ | $c_{ETAS}$ | $b_{ETAS}$ | std $b$ | $\alpha$ | std $\alpha$ | $a_1$ |
|---|---|---|---|---|---|---|---|---|---|---|---|---|---|---|---|
| A06 | 06-05-1976 | 6.1 | 2.4 | 509 | 0.87 | 0.02 | -1.57 | 0.82 | 1.07 | 0.03 | 0.85 | 0.07 | 0.73 | 0.05 | -1.32 |
| B14 | 19-09-1979 | 5.0 | 2.0 | 245 | | | | | 1.02 | 0.02 | 1.63 | 0.20 | 0.69 | 0.07 | 1.08 |
| B17 | 29-04-1984 | 5.2 | 2.0 | 244 | 1.27 | 0.39 | -1.34 | 1.04 | 1.28 | 0.17 | 1.04 | 0.13 | 1.10 | 0.15 | -1.96 |
| A13 | 07-05-1984 | 5.8 | 2.0 | 371 | | | | | 1.14 | 0.04 | 0.90 | 0.11 | 0.64 | 0.05 | -0.80 |
| B35 | 26-06-1993 | 4.3 | 1.3 | 488 | 0.78 | 0.10 | -1.06 | 0.91 | 0.94 | 0.02 | 0.91 | 0.12 | 1.00 | 0.10 | -2.02 |
| A30 | 15-10-1996 | 5.5 | 1.7 | 207 | 0.93 | 0.08 | -1.98 | 0.98 | 1.00 | 0.01 | 0.97 | 0.11 | 0.77 | 0.07 | -1.29 |
| | 26-09-1997 | 5.8 | 1.8 | 3063 | | | | | 1.29 | 0.06 | 0.89 | 0.03 | 0.62 | 0.02 | -0.72 |

# Appendix A

## Dataset 1 (Italy, Lolli and Gasperini, 2003)

| Seq. | $M_m$ | $M_{min}$ | p | c | b | a | N |
|---|---|---|---|---|---|---|---|
| B01 | 4.7 | 1.5 | 2.07 | 0.07 | 0.88 | -2.36 | 29 |
| B02 | 5.3 | 2.3 | 1.26 | 0.06 | 0.98 | -1.74 | 47 |
| B05 | 5.4 | 2.6 | 0.64 | 0.03 | 1.14 | -2.18 | 42 |
| B07 | 6.1 | 2.4 | 0.91 | 0.02 | 0.82 | -1.62 | 539 |
| B08 | 5.5 | 2.4 | 0.68 | 0.04 | 1.03 | -1.71 | 52 |
| B10 | 5.7 | 2.4 | 1.01 | 0.23 | 0.94 | -1.59 | 179 |
| A07 | 4.7 | 2.2 | 0.73 | 1.00 | 1.13 | -2.04 | 22 |
| B12 | 5.0 | 2.1 | 0.64 | 0.10 | 1.05 | -2.01 | 57 |
| B15 | 4.3 | 1.4 | 1.09 | 0.34 | 0.94 | -0.94 | 131 |
| B16 | 6.5 | 2.6 | 0.92 | 0.44 | 0.97 | -2.15 | 133 |
| B17 | 5.2 | 2.0 | 1.27 | 0.39 | 1.04 | -1.34 | 297 |
| B18 | 5.8 | 2.2 | 0.86 | 0.01 | 0.79 | -1.69 | 402 |
| B19 | 5.3 | 2.0 | 1.27 | 0.26 | 0.84 | -0.94 | 328 |
| B20 | 4.4 | 2.0 | 0.88 | 4.16 | 1.22 | -1.23 | 86 |
| A15 | 4.4 | 1.9 | 1.36 | 0.52 | 0.79 | -0.68 | 72 |
| B23 | 4.6 | 1.9 | 0.78 | 0.06 | 0.74 | -1.63 | 24 |
| B28 | 5.5 | 1.9 | 1.06 | 0.07 | 0.94 | -1.91 | 156 |
| B29 | 5.1 | 1.9 | 1.93 | 7.45 | 1.04 | -0.86 | 20 |
| A19 | 4.3 | 1.7 | 0.68 | 0.44 | 0.91 | -1.96 | 19 |
| B33 | 4.3 | 1.7 | 1.01 | 2.14 | 0.97 | -1.56 | 22 |
| B34 | 4.5 | 1.4 | 0.62 | 0.01 | 1.48 | -3.34 | 118 |
| B35 | 4.3 | 1.3 | 0.78 | 0.10 | 0.91 | -1.06 | 547 |
| A22 | 4.4 | 1.2 | 0.85 | 0.01 | 0.99 | -2.60 | 19 |
| A23 | 4.4 | 1.8 | 0.70 | 0.03 | 0.94 | -2.41 | 20 |
| B37 | 4.7 | 2.2 | 1.14 | 0.12 | 1.10 | -1.78 | 24 |
| A24 | 4.8 | 1.7 | 0.80 | 0.11 | 0.80 | -1.25 | 127 |
| A25 | 4.3 | 1.4 | 0.95 | 0.01 | 1.31 | -2.88 | 33 |
| B40 | 4.8 | 1.4 | 0.97 | 0.07 | 1.26 | -3.43 | 44 |
| A28 | 4.4 | 1.2 | 0.93 | 0.01 | 0.83 | -2.02 | 39 |
| A30 | 5.5 | 1.7 | 0.93 | 0.08 | 0.98 | -1.98 | 237 |

**Dataset 2 (New Zealand, Eberhart-Phillips, 1998)**

| Seq. | $M_m$ | $M_{min}$ | p | c | b | a | N |
|---|---|---|---|---|---|---|---|
| A | 6.6 | 3.0 | 1.61 | 0.22 | 1.03 | -1.65 | 421 |
| B | 6.7 | 2.8 | 1.23 | 0.06 | 0.86 | -1.95 | 153 |
| C | 5.9 | 2.6 | 1.05 | 0.04 | 0.78 | -0.90 | 341 |
| D | 6.3 | 2.7 | 0.95 | 0.03 | 1.27 | -3.01 | 267 |
| E | 6.4 | 2.7 | 1.25 | 0.06 | 0.82 | -1.27 | 364 |
| F | 5.5 | 2.2 | 1.13 | 0.34 | 0.87 | -1.05 | 269 |
| G | 6.0 | 2.5 | 0.87 | 0.01 | 0.88 | -1.66 | 196 |
| H | 5.5 | 2.2 | 0.90 | 0.01 | 0.98 | -1.92 | 152 |
| I | 5.5 | 2.9 | 0.82 | 0.01 | 0.90 | -0.83 | 221 |
| J | 5.5 | 2.4 | 0.98 | 0.01 | 0.94 | -1.45 | 214 |
| K | 5.7 | 2.7 | 0.74 | 0.01 | 1.06 | -1.84 | 154 |
| L | 6.3 | 3.2 | 1.38 | 0.09 | 1.09 | -1.66 | 276 |
| M | 5.7 | 2.4 | 0.20 | 0.02 | 1.15 | -3.34 | 25 |
| N | 6.2 | 3.0 | 1.01 | 0.01 | 1.27 | -3.03 | 75 |
| O | 6.7 | 3.3 | 1.21 | 0.07 | 1.10 | -1.79 | 579 |
| P | 7.1 | 4.6 | 1.38 | 0.33 | 1.16 | -1.04 | 250 |
| Q | 6.2 | 2.5 | 1.02 | 0.01 | 1.04 | -2.07 | 468 |

**Dataset 3 (California, Reasenberg and Jones, 1989; Jones personal communication)**

| Seq. | $M_m$ | $M_{min}$ | p | c | b | a | N |
|---|---|---|---|---|---|---|---|
| C1 | 6.3 | 3.3 | 1.37 | 0.05 | 1.00 | -0.95 | 570 |
| C2 | 5.1 | 2.1 | 1.70 | 0.05 | 0.56 | -1.34 | 21 |
| C3 | 6.0 | 3.0 | 1.12 | 0.05 | 0.94 | -2.12 | 42 |
| C4 | 5.4 | 2.4 | 0.59 | 0.05 | 0.87 | -0.86 | 62 |
| C5 | 6.9 | 3.9 | 1.35 | 0.05 | 0.58 | -1.24 | 39 |
| C6 | 5.9 | 2.9 | 0.96 | 0.05 | 0.94 | -1.83 | 71 |
| C7 | 6.5 | 3.5 | 0.67 | 0.05 | 0.91 | -2.36 | 103 |
| C8 | 6.3 | 3.3 | 1.30 | 0.05 | 0.89 | -1.10 | 185 |
| C9 | 6.2 | 3.2 | 1.11 | 0.05 | 0.90 | -1.66 | 149 |
| C10 | 5.5 | 2.5 | 1.06 | 0.05 | 0.67 | -0.75 | 82 |
| C11 | 6.0 | 3.0 | 1.06 | 0.05 | 0.91 | -1.77 | 101 |
| C12 | 5.9 | 2.9 | 0.94 | 0.05 | 1.27 | -2.29 | 112 |
| C13 | 5.7 | 2.7 | 1.11 | 0.05 | 0.85 | -1.39 | 41 |
| C14 | 5.6 | 2.6 | 1.18 | 0.05 | 0.84 | -1.21 | 213 |
| C15 | 5.0 | 2.0 | 1.02 | 0.05 | 0.66 | -1.12 | 18 |
| C16 | 7.7 | 4.7 | 1.05 | 0.05 | 0.94 | -2.11 | 1216 |
| C17 | 6.2 | 3.2 | 1.20 | 0.05 | 0.92 | -1.67 | 113 |
| C18 | 6.8 | 3.8 | 0.87 | 0.05 | 0.88 | -1.49 | 69 |
| C19 | 7.2 | 4.2 | 1.11 | 0.05 | 0.61 | -1.33 | 20 |
| C20 | 5.2 | 2.2 | 2.05 | 0.05 | 0.56 | -1.20 | 18 |
| C21 | 5.3 | 2.3 | 1.08 | 0.05 | 0.83 | -1.38 | 188 |
| C22 | 5.8 | 2.8 | 0.80 | 0.05 | 0.74 | -1.59 | 89 |
| C23 | 6.0 | 3.0 | 1.06 | 0.05 | 0.73 | -1.28 | 40 |
| C24 | 6.5 | 3.5 | 1.15 | 0.05 | 0.99 | -2.24 | 136 |
| C25 | 5.2 | 2.2 | 1.04 | 0.05 | 0.93 | -1.26 | 37 |
| C26 | 6.6 | 3.6 | 1.24 | 0.05 | 1.04 | -2.06 | 627 |
| C27 | 5.0 | 2.0 | 0.79 | 0.05 | 1.01 | -1.70 | 1247 |
| C28 | 5.5 | 2.5 | 0.75 | 0.05 | 0.57 | -1.32 | 24 |
| C29 | 5.2 | 2.2 | 0.79 | 0.05 | 0.94 | -2.27 | 79 |
| C30 | 5.2 | 2.2 | 1.07 | 0.05 | 1.00 | -2.15 | 105 |
| C31 | 5.7 | 2.7 | 0.93 | 0.05 | 0.59 | -0.78 | 244 |
| C32 | 5.0 | 2.0 | 0.56 | 0.05 | 0.81 | -1.40 | 110 |
| C33 | 5.2 | 2.2 | 0.82 | 0.05 | 1.06 | -2.45 | 215 |
| C34 | 5.1 | 2.1 | 1.12 | 0.05 | 1.20 | -2.22 | 298 |
| C35 | 5.8 | 2.8 | 1.38 | 0.05 | 0.80 | -1.44 | 180 |
| C36 | 5.0 | 2.0 | 1.17 | 0.05 | 0.87 | -1.33 | 185 |
| C37 | 5.2 | 2.2 | 1.01 | 0.05 | 0.99 | -1.39 | 648 |
| C38 | 5.9 | 2.9 | 1.04 | 0.05 | 0.95 | -2.55 | 96 |
| C39 | 6.6 | 3.6 | 1.49 | 0.05 | 1.00 | -2.01 | 1273 |
| C40 | 5.9 | 2.9 | 1.08 | 0.05 | 0.90 | -2.17 | 159 |
| C41 | 5.5 | 2.5 | 1.10 | 0.05 | 0.95 | -2.39 | 82 |
| C42 | 6.5 | 3.5 | 0.93 | 0.05 | 0.93 | -1.39 | 1887 |
| C43 | 6.1 | 3.1 | 1.16 | 0.05 | 0.98 | -2.23 | 338 |
| C44 | 5.7 | 2.7 | 0.96 | 0.05 | 0.90 | -1.67 | 221 |
| C45 | 5.7 | 2.7 | 1.34 | 0.05 | 0.99 | -2.07 | 255 |
| C46 | 5.3 | 2.3 | 0.90 | 0.05 | 0.77 | -1.46 | 182 |
| C47 | 5.4 | 2.4 | 0.95 | 0.05 | 0.79 | -1.83 | 40 |
| C48 | 5.7 | 2.7 | 1.08 | 0.05 | 0.89 | -1.19 | 242 |
| C49 | 6.5 | 3.5 | 1.06 | 0.05 | 0.89 | -1.47 | 2789 |
| C50 | 5.2 | 2.2 | 1.32 | 0.05 | 0.50 | -1.22 | 15 |
| C51 | 6.2 | 3.2 | 0.87 | 0.05 | 1.01 | -3.25 | 141 |
| C52 | 6.2 | 3.2 | 1.08 | 0.05 | 0.96 | -1.71 | 1067 |
| C53 | 5.8 | 2.8 | 0.94 | 0.05 | 1.04 | -2.41 | 235 |
| C54 | 5.7 | 2.7 | 1.08 | 0.05 | 1.10 | -2.93 | 73 |
| C55 | 5.7 | 2.7 | 0.91 | 0.05 | 1.15 | -3.14 | 108 |
| C56 | 5.9 | 2.9 | 1.17 | 0.05 | 0.93 | -1.59 | 1436 |
| C57 | 5.5 | 2.5 | 0.75 | 0.05 | 1.34 | -2.58 | 1639 |
| C58 | 6.5 | 3.5 | 1.19 | 0.05 | 0.92 | -1.77 | 776 |
| C59 | 5.4 | 2.4 | 1.30 | 0.05 | 1.20 | -2.60 | 111 |
| C60 | 5.5 | 2.5 | 1.20 | 0.05 | 1.02 | -2.30 | 94 |
| C61 | 6.0 | 3.0 | 1.51 | 0.05 | 0.67 | -1.41 | 183 |
| C62 | 6.5 | 3.5 | 1.30 | 0.05 | 0.86 | -1.97 | 752 |
| C63 | 5.4 | 2.4 | 1.22 | 0.05 | 1.07 | -2.90 | 95 |
| C64 | 4.9 | 1.9 | 1.44 | 0.05 | 1.19 | -3.53 | 15 |